\documentclass[review]{elsarticle}
\usepackage[utf8]{inputenc}
\usepackage{hyperref}
\usepackage{lineno}
\usepackage{amsmath}
\usepackage{fixmath}
\usepackage{amsfonts}
\usepackage{amssymb}
\usepackage{array}
\usepackage{float}
\usepackage{graphicx}
\usepackage{subcaption}
\usepackage{booktabs}
\usepackage[a4paper, total={6.6in, 10.4in}]{geometry}%
\usepackage{lineno,hyperref}
\usepackage{multirow}
\usepackage{xcolor}
\usepackage{color}
\usepackage{appendix}
\usepackage{gensymb}
\usepackage{makecell}
\modulolinenumbers[1]

\journal{none}

\begin{document}

\begin{frontmatter}

\title{A unified framework for prediction of vortex-induced vibration based on the nonlinear data-driven identification of general wake oscillator modeling}


\author[address0,address1]{Zhi Cheng\corref{mycorrespondingauthor}}
\ead{vamoschengzhi@gmail.com}
\cortext[mycorrespondingauthor]{Corresponding author}

\author[address2]{Fue-Sang Lien}



\author[address1]{Earl H. Dowell}

\address[address0]{Mechanical Engineering, The University of British Columbia, 2329 West Mall, Vancouver, British Columbia, V6T 1Z4, CANADA.}

\address[address1]{Pratt School of Engineering, Duke University, 2080 Duke University Road, Durham, North Carolina, 27708, USA.}

\address[address2]{Mechanical and Mechatronics Engineering, University of Waterloo, 200 University Avenue West, Waterloo, Ontario, N2L 3G1, CANADA.}


\begin{abstract}

In this paper, we present novel identification strategies to develop a unified framework for vortex-induced vibration (VIV) prediction based on the general semi-empirical wake oscillator. Greybox nonlinear system identification method accompanying high-fidelity computational fluid dynamics (CFD) and/or experimental data could be applied for the identification process. 
The proposed template of general wake oscillators contains low- to high-order damping terms to be identified for characterizing the possible flow dynamics.
Two different strategies, including individual identification of single wake oscillator and overall identification of coupled VIV control equations, are proposed. 
VIV system consisting of an elastically-mounted circular cylinder submerged in laminar flow at Reynold number ($Re$) of 100 is considered. Both two strategies are tested and exhibit high accuracy. 
The second strategy, i.e., overall identification of coupled VIV control equations, would be more suitable for the future framework owing that its training process considers the effect of fluid damping.
A detailed mathematical introduction to future works on framework development covering the wide $Re$ range is addressed. The proposed unified framework is a landmark update of past wake oscillators both in terms of prediction accuracy and physical principles and has considerable research significance and practical engineering value.

\end{abstract}


\end{frontmatter}



\section{Introduction}

Flow-induced vibration (FIV) is a prevalent fluid-solid coupling behavior both within nature and engineering systems \cite{JOHNS1998315, ZhangBo2016, Billah1991}. 
In terms of underlying mechanisms, FIV can be classified as flutter lock-in, resonance lock-in, galloping, surge, etc. \cite{Modi1998, Waals2007, SurenChen2004}. Among those served possibilities, the resonance lock-in, i.e., vortex-induced vibration (VIV), is one of the most widely studied behaviors. 
Alternate shedding vortexes will be formed downstream when the cylinder is submerged in air or water flow, and those vortics will push the cylinder to produce periodic vibration. 
The moving cylinder under pulsating lift will, in turn, have an impact on the surrounding flow field. When the vortex-shedding frequency is close to the structural natural frequency, the system will produce lock-in behavior, which will generate a significantly amplified structural amplitude that can damage the structure \cite{ChengEtal2022}. The VIV problem exists over a wide range of practical engineering designs, such as bridges, marine structures, marine cables, anchor wave systems, aircraft, etc. Therefore, the VIV problem has always been a matter of interest in industrial design and also academic research.

There are several methodologies applied to the investigation and prediction of VIV, among which computational fluid dynamics (CFD) is widely used and quite accurate \cite{vaziri_ekmekci_2022,kovalev_eshbal_vanhout_2022,bourguet_karniadakis_triantafyllou_2013, paidoussis2010fluid}. However, the conventional CFD method is often too expensive for engineering applications. 
In recent years, machine learning tools have been introduced into the research on flow-induced vibration \cite{BAI2022105266, Chizfahm2022, KIM2022114551}.
Despite the buzz around machine learning today, the complexity of the machine learning process, the near-black-box training results, and the difficulty in implementation of predictive models have made it unfriendly and even inaccessible for industry applications. In addition, the models obtained from machine learning do not have a clear physical meaning, making their robustness and reliability somewhat questionable.
In contrast, the semi-empirical modeling method is efficient/convenient, physically-meaningful, and understandable. A typical semi-empirical modeling method for investigating the VIV problem is the wake oscillator model. 
The wake oscillator model can be applied to characterize the fluid dynamics, i.e., the wake of the fluid is regarded as a nonlinear oscillator. 
In view of the whole VIV system, the mutual coupling between the wake oscillator and the structural model could thereby predict the structural displacement response as well as the variation of the lift fluid force during the development of VIV.

Various wake oscillator models have been introduced in past studies to describe VIV and associated lock-in phenomena. The prevalent models describing fluid oscillation behaviors are Van Der Pol, Rayleigh, and other classical models \cite{nayfeh2011introduction, FACCHINETTI2004123, 8084644}.
More specifically, the most widely-used and fundamental wake oscillator originates from the Van Der Pol equation can be expressed as:
\begin{linenomath}
\begin{equation}\label{structuraleqa5}
\ddot{q}+\varepsilon \varOmega _f( q^2-1 ) \dot{q}+\varOmega _{f}^{2}q=F
\end{equation}
\end{linenomath}
where $F$ represents the effect acted by structural vibration on the wake dynamics, $q$ = $2C_L/C_{L,0}$ is a non-dimensional flow variable, $C_{L,0}$ is lift coefficients' amplitude for a structure/body at rest and $C_{L}$ is the flow lift. The Van Der Pol form contains the 3rd-order nonlinear damping terms, whereas another kind of wake oscillator in Rayleigh form is expressed as:
\begin{linenomath}
\begin{equation}\label{Rayleighstructuraleqa5}
\ddot{q}+\varepsilon \varOmega _f( {\dot q}^2-1 ) \dot{q}+\varOmega _{f}^{2}q=F
\end{equation}
\end{linenomath}
in which the modifications are made for the internal damping terms (compared to Van Der Pol) in order to be compatible with some specific scenarios.
The structural governing equations are well established, hence the model improvement should be mainly concentrated on how the internal damping term is adjusted inside the wake oscillator.
Ogink and Metrikine \cite{OGINK20105452} tried to enhance the wake oscillator by introducing frequency-dependent coupling. The coefficient of the damping term within the control equation is no longer a fixed value, but a semi-empirical value that varies with the vortex-shedding frequency.
However, this work also further suggested that it is impossible to ﬁnd one set of frequency-dependent coefﬁcients that conform to the experimental data at all amplitudes of cylinder motion if the wake oscillator is modeled with a Van der Pol equation.
Farshidianfar and Dolatabadi \cite{Farshidianfar2013} introduced a novel wake oscillator modified on the base of original Van Der Pol equation (cf. with Eq. \ref{structuraleqa5}) and a ﬁfth-order aerodynamic damping term is added:
\begin{linenomath}
\begin{equation}\label{forthstructuraleqa5}
\ddot{q}+\varepsilon \varOmega _f(1 - \beta q^2 + \lambda q^4) \dot{q}+\varOmega _{f}^{2}q=F ,
\end{equation}
\end{linenomath}
in addition, the control coefficients $\beta$ and $\lambda$ of the damping terms in the model are variable. This work shows that such improvements could improve the prediction accuracy of the model for specific experimental data.


In addition to VIV response in the transverse direction, some studies have considered flow-induced vibration in the streamline direction, and thus two sets of coupled fluid-structure equations are defined for each of the two directions, respectively. 
Srinil and Zanganeh \cite{SRINIL201283} proposed adding a higher-order damping model of different orders to the wake oscillators in each direction to improve the prediction accuracy.
Qu and Metrikine \cite{QU2020106732} proposed a fluid model with only one single wake oscillator formulation that enables coupling with the structural equations in both the transverse and in-line directions.
Nevertheless, the larger structural amplitude of transverse vibration compared to in-line vibration is still the primary focus of past research, and thus this paper focuses on the transverse VIV response.

In addition to tuning the damping term inside wake oscillators, the force term (i.e., $F$) at the right end of the above wake oscillator formulations (viz., equations \ref{structuraleqa5},\ref{Rayleighstructuraleqa5},\ref{forthstructuraleqa5}) also involves a relevant coupling of structural moving information (including three terms: structural displacement, velocity, and acceleration).
Hartlen and Currie \cite{Hartlen1970} first applied the velocity as the coupled source term. Facchinetti and Langre \cite{FACCHINETTI2004123} also suggested using displacement and acceleration as source terms. 
Past works adjusted the fluid-induced damping term in the structural equation of motions to get the optimal conformance between semi-empirical models (via wake oscillators) and measurement data \cite{gao2018novel}.
In addition to local adjustments, regarding the overall form of the structural governing equations, Farshidianfar and Zanganeh \cite{FARSHIDIANFAR2010430} have attempted to improve the prediction accuracy by writing the structural equations in the form of a Van Der Pol oscillator, similar to the wake oscillator. However, instead of adjusting structural control equations, it makes more physical sense to evaluate the degree of nonlinearity of the wake oscillator and optimize different order terms to characterize the fluid properties.

Despite the valuable past developments on the wake oscillator, all the past works focus on partial modifications and step-by-step improvements to facilitate the matching of prediction and specific experimental data.
For the purpose of making model prediction fully compatible with high-fidelity CFD and/or experimental results,
we will design one general wake oscillator template in which various damping terms (from low- to high-orders) and three structural motion terms (including displacements, velocities, and accelerations) are considered. It is noted that the coefficients corresponding to each term are pending at this stage.
This specific wake oscillator is thereby determined using high-fidelity experimental/numerical data based on certain identification tools.
There is a dearth of past research on wake oscillators using the above strategy. 
Xu and Ge \cite{XU2015192} used a genetic algorithm to optimize the model of the wake oscillator, allowing it to simulate the experimental results as closely as possible. The predicted results are consistent with the experimental result, but only for one single/specific scenario.

Following Dowell \cite{DOWELL1981251}, we introduce strategies for developing a comprehensive prediction framework for the common range of Reynolds numbers as well as reduced velocities. The identification process will be implemented via the grey-box dynamical identification models.
This work is arranged as follows:
the numerical methods including the identification tool adopted in the present study are introduced in Section 2. 
Two identification strategies are introduced and model validation for the computational fluid dynamics (CFD) method is also conducted in this section.
Section 3 introduces the implementation process of strategy 1, i.e., direct identification of single wake oscillator, with the help of the CFD calculations on the forced oscillation of a circular cylinder.
In Section 4, we discuss strategy 2 which regards the coupled VIV control equations (including both wake oscillator and structural equations) as a whole. Two scenarios of self-excited vibration are considered here as examples.
At the end of Section 4, the concept of establishing an overall prediction framework is described from a mathematical point of view.
Finally, the conclusion is presented in Section 5.

\section{Model methodology}

\subsection{Prediction model based on a universal wake oscillator}

The structural equations for the transverse motion of an elastically supported rigid slender model (with a single-degree-of-freedom) submerged in the fluid can be expressed as:
\begin{linenomath}
\begin{equation}\label{structuraleqa1}
m\ddot{Y}+c\dot{Y}+kY=W
\end{equation}
\end{linenomath}
where $m$ is mass, $c$ is the structural damping, $k$ is the spring stiffness, $Y$ is the structural transverse displacements, $W$ is the fluid forces acted on the body surface. Mass $m$ includes the structural mass $m_s$ and flow added mass $m_f$, which could be represented as:
\begin{linenomath}
\begin{equation}\label{structuraleqa2}
m = m_s+m_f, m_f=C_M\rho D^2\pi/4,\mu=m/(\rho D^2)
\end{equation}
\end{linenomath}
where $\rho$ is the fluid density, $\mu$ is the non-dimensional mass ratio, $C_M$ is the added mass coefficients, and $D$ is the characteristic length.
Damping $c$ is the sum of the structural damping $c_s$ and the fluid damping $c_f$:
\begin{linenomath}
\begin{equation}\label{structuraleqa3}
c = c_s+c_f,
\end{equation}
\end{linenomath}
where structural damping $c_s=2m\varOmega _s\zeta$, $\zeta$ is the structural damping ratio and $\varOmega _s$ (=$\sqrt{k/m}$) is the structural natural frequency. $c_f$ (=$\gamma \varOmega \rho D^2$) is the fluid added damping, where $\varOmega$ is the structural vibration frequency, $\gamma$ is the stall parameter, a function of structural displacements and correlated to drag coefficients $C_D$, which could be represented as:
\begin{linenomath}
\begin{equation}\label{structuraleqa4}
\gamma =\frac{C_{D}}{4\pi S_t}.
\end{equation}
\end{linenomath}

The fluid force $W$ acted on the structure is:
\begin{linenomath}
\begin{equation}\label{structuraleqa4}
S=\frac{1}{2}\rho U^2DC_L
\end{equation}
\end{linenomath}
where $C_L$ is the lift coefficient.

The dynamic characteristics of the vortex-shedding behaviors could be represented using a nonlinear wake oscillator model.
With respect to the wake oscillator in Van Der Pol form, the coupling of Eq. \ref{structuraleqa1} and Eq. \ref{structuraleqa5} is applied to the VIV system. Introducing the non-dimensional time $t = T \varOmega_f$ and the non-dimensional displacement $y = Y/D$, Eqs. \ref{structuraleqa1} and \ref{structuraleqa5} are transformed into the dimensionless form as:
\begin{linenomath}
\begin{equation}\label{structuraleqa6}
\ddot{y}+( 2\zeta \delta +\gamma /\mu ) \dot{y}+\delta ^2y=w,
\end{equation}
\end{linenomath}
\begin{linenomath}
\begin{equation}\label{structuraleqa7}
\ddot{q}+\varepsilon ( q^2-1 ) \dot{q}+q=f,
\end{equation}
\end{linenomath}
where $\delta$ is the reduced structural natural frequency:
\begin{linenomath}
\begin{equation}\label{structuraleqa8}
\delta =\frac{\varOmega _s}{\varOmega _f}=\frac{\varOmega _s}{2\pi S_tU/D}=\frac{1}{S_tU_r},U_r=\frac{2\pi U}{\varOmega _sD}.
\end{equation}
\end{linenomath}

The non-dimensional fluid force $w$ acted on structure is:
\begin{linenomath}
\begin{equation}\label{structuraleqa9}
w=Mq,M=\frac{C_{L}^{0}}{16\pi ^2S_t^2\mu}.
\end{equation}
\end{linenomath}

With respect to the development of the wake oscillator, past work has focused on determining and modifying source term $f$ to achieve optimal prediction accuracy.
As stated in Introduction section, unlike past work that dowelled on the terms' modification with varied order on the left side and the addition/subtraction of the source terms on the right side of Eq. \ref{structuraleqa7}, this paper constructs for the first time a more generalized template of the wake oscillator, and uses the model identification tool (described later) to determine the parameters within the template based on the VIV response data provided by the experimental or higher-order CFD model. As an alternative to Eq. \ref{structuraleqa7}, the general wake oscillator template in present work is denoted as:
\begin{linenomath}
\begin{equation}\label{structuraleqa10}
\ddot{q}+q+( \varepsilon _{10}+\varepsilon _{12}q^2+\varepsilon _{14}q^4+\varepsilon _{16}q^6 ) \dot{q}+( \varepsilon _{20} ) \dot{q}^2+( \varepsilon _{30} ) \dot{q}^3 = A_{F0}y+A_{F1}\dot{y}+A_{F2}\ddot{y}.
\end{equation}
\end{linenomath}

The parameters, $\varepsilon _{10}$, $\varepsilon _{12}$, $\varepsilon _{14}$, $\varepsilon _{16}$, $A_{F0}$, $A_{F1}$, will be correlated to the properties (including Reynolds number, mass ratio, reduced velocity, etc.) of the VIV system.



\subsection{Nonlinear system identification}


Greybox Nonlinear System Identification method \cite{WERNHOLT2005356, LJUNG2004399} will be used for the identification process herein for the wake oscillator model.
The model will be described by a continuous-time state-space structure \cite{Gunnar1641970, WERNHOLT2005356, LJUNG2004399}.

\begin{linenomath}
\begin{equation}\label{wakeoscillatorstatusequation}
\begin{aligned}
s( t ) &=f( t,s( t ) ,\theta ,u( t ) ),
\\
o( t ) &=h( t,s( t ) ,\theta ,u( t ) ) +e( t ),
\end{aligned}
\end{equation}
\end{linenomath}
where both $f$ and $h$ are nonlinear functions, $s(t)$ is the state vector, $u(t)$ are the input signal and $o(t)$ are the output signal.
$e(t)$ is white measurement noise and $\theta$ is the unknown parameter vector. The aim of the identification is to find the parameters, that
given data from the physical system, will minimize the criterion
\begin{linenomath}
\begin{equation}\label{wakeoscillatorstatusequation}
\begin{aligned}
V( \theta ) =\frac{1}{N}\sum_{t=1}^N{\epsilon ^2}( t,\theta )
\end{aligned}
\end{equation}
\end{linenomath}
where ${\epsilon}( t,\theta )$ is the prediction error ${\epsilon}( t,\theta )$ = $o(t)-\hat{o}( t,\theta )$.
The prediction $\hat{o}( t,\theta )$ then becomes the simulated output of the concerned model with the input $u(t)$ (without $e(t)$).
The toolbox estimates the parameter vector $\theta$ by applying a prediction error method, which performs a numerical optimization of
the criterion. The numerical optimization is performed by an iterative numerical search algorithm.

In the present work, we pursue conduct two identification strategies including: (1) the direct identification of wake oscillator (viz., Eq. \ref{structuraleqa10}); and (2) the identification of the coupled VIV models (viz., Eqs. \ref{structuraleqa10} and \ref{structuraleqa6}).
With respect to strategy 1, the structural motion information including displacements $y$, velocity $\dot{y}$, acceleration $\ddot{y}$ are regarded as input and dynamical coefficients $q$ and $\dot{q}$ are output. Equation \ref{structuraleqa6} could be rewritten with state vector as $[ q,\dot{q} [ ^T =[ s_1 ,s_2 ]$ and dynamical model as:
\begin{linenomath}
\begin{equation}\label{wostatusequation}
\begin{aligned}
ds_1 &=s_2 ,
\\
ds_2 &=A_{F0}y+A_{F1}\dot{y}+A_{F2}\ddot{y}-( s_1+( \varepsilon _{10}+\varepsilon _{12}s_1^2+\varepsilon _{14}s_1^4+\varepsilon _{16}s_1^6 ) s_2+( \varepsilon _{20} ) s_2^2+( \varepsilon _{30} ) s_2^3 ).
\end{aligned}
\end{equation}
\end{linenomath}

To obtain the training data for identification process of strategy 1, the present work will perform a high-fidelity CFD calculation for the forced oscillation of the circular cylinder. More specifically, we will design the motion/oscillation patterns (i.e., input) of the body and obtain the dynamical response such as force coefficients (i.e., output) at the specific Reynolds number $Re$, and then introduce those input and output data as the training data-set for the identification. For strategy 2, the coupling Eqs. \ref{structuraleqa10} and \ref{structuraleqa6} will be derived with state vector expressed as $[ q,\dot{q},y,\dot{y} ] ^T = [ s_1 ,s_2 ,s_3 ,s_4 ]$ and dynamical model expressed as:
\begin{linenomath}
\begin{equation}\label{VIVstatusequation}
\begin{aligned}
ds_1 =& s_2,
\\
ds_2 =& A_{F0}s_3+A_{F1}s_4+A_{F2}( \frac{2}{\pi}( \frac{U_0}{D} ) ^2\frac{C_L}{m*}-4\pi \zeta f_ns_4-( 2\pi f_n ) ^2s_3 ) \\
&-( s_1+( \varepsilon _{10}+\varepsilon _{12}s_1^2+\varepsilon _{14}s_1^4+\varepsilon _{16}s_1^6 ) s_2+( \varepsilon _{20} ) s_2^2+( \varepsilon _{30} ) s_2^3 ),
\\
ds_3 =& s_4,
\\
ds_4 =& w-( 2\varepsilon_{fn} \zeta \delta +\gamma /\mu ) s_4-\delta ^2s_3,
\end{aligned}
\end{equation}
\end{linenomath}
where there is no input and/or output for this dynamic model. For the parameters identification of the coupled dynamic model, in this work we first calculate the dynamic response of one VIV system (at a specific combination of ($Re, m^*, U_r$)) based on the high-fidelity CFD method, and then we use the structural motion data and the dynamical coefficients as the training basis for Equation \ref{VIVstatusequation}.
For the processing of strategy 2, the fluid damping term is deleted from the structural control equation because the identification process of the overall coupled VIV modeling already considers the potential nonlinear effect of fluid damping on the structural motion. 
Moreover, it is observed that one additional coefficient $\varepsilon_{fn}$ is added inside the structural equation (or, last line) in Eq. \ref{VIVstatusequation}. 
This tuning parameter is added to correct the difference between the actual structural vibration frequency and the structural natural frequency due to the effect of added fluid mass \cite{wang2021illuminating}.

\subsection{Full-order model of computational fluid dynamics and its validation}
\label{CFD_m}

The fluid-solid interaction problem concerned in the present work consists of a square column elastically mounted on a linear spring and immersed in a three-dimensional uniform fluid flow. For resolving the response of VIV system, the continuity and momentum transport equations and the structural equation are applied to govern the fluid flow and the motion of the elastically-supported bluff body. Specifically, the continuity and momentum transport equations are:
\begin{linenomath}
	\begin{equation}\label{eq:ns1}
		\frac{\partial u_i}{\partial x_i}=0\ ,
	\end{equation}
\end{linenomath}
and
\begin{linenomath}
	\begin{equation}\label{eq:ns2}
		\frac{\partial u_i}{\partial t}+( u_j-\tilde{u}_j \delta_{ij} ) \frac{\partial u_i}{\partial x_j}=-\frac{1}{\rho}\frac{\partial p}{\partial x_i}+\nu \frac{\partial ^2u_i}{\partial x_j\partial x_j}\ .
	\end{equation}
\end{linenomath}

The structural equation for the motion of the bluff body in CFD method is modified as:
\begin{linenomath}
	\begin{equation}\label{CFDeq:structMod}
	  m\ddot{Y}+c_s\dot{Y}+kY=W
	\end{equation}
\end{linenomath}
where $c_f$ is deleted from $c$ (compared to eq. \ref{structuraleqa1}) owing that the effect of fluid damping will be considered in the high-fidelity CFD resolution via Eqs. \ref{eq:ns1}, \ref{eq:ns2}.

In equations (\ref{eq:ns1}) and (\ref{eq:ns2}), \(x_i\) is $i$-th component of a Cartesian coordinate vector $\vec x$ with $i \equiv$ 1, 2, and 3 corresponding to the streamwise $x$-, transverse $y$-, and spanwise $z$-directions, respectively; \(p\) is the pressure; $\delta_{ij}$ is the Kronecker delta function; \(t\) is the time; $\rho$ and $\nu$ represent the density and kinematic viscosity of the fluid, respectively; \(u_i\) is the $i$-th component of fluid velocity; and, $\tilde{u}_j \equiv dx_j/dt$ represents $j$-th component of the grid velocity arising from the motion of the body (structure) immersed in the flow.

The CFD open-source software OpenFOAM \cite{openfoamv2006} is used for the calculations in this paper. 
Consistent with the methodology of our previous works in aeroacoustics \cite{CHENG2023652}, the Navier-Stokes equations are discretized by the finite volume method. The transient terms are discretized using the second-order implicit Eulerian scheme, and the advection, pressure gradient and diffusion terms are discretized using the second-order Gaussian integration scheme.
An adaptive time-step $\Delta t$ technique is used to ensure that the maximum Courant-Friedrichs-Lewy (CFL) number ${\rm CFL}_{\rm max}$ is limited to 0.8 (${\rm CFL}_{\rm max} \equiv{\|\vec u \|\Delta t}/{\Delta x_{\rm min}}$, 
$\|\vec u\|$ is the magnitude of the fluid velocity $\vec u$ and $\Delta x_{\rm min}$ is the size of the smallest grid cell in the computational domain). 
An explicit second-order symplectic method \citep{Dullweber1997} is applied to integrate the structural equations of motion.
The weakly coupled approach \cite{WANG201912} is applied to solve the fluid-structure interaction that links the fluid flow equations (\ref{eq:ns1}, \ref{eq:ns2}) with the structural equation of motion (\ref{CFDeq:structMod}).

\begin{figure}[t]
	\centering
	\begin{subfigure}{0.4655\linewidth}
		\includegraphics[width=\linewidth]{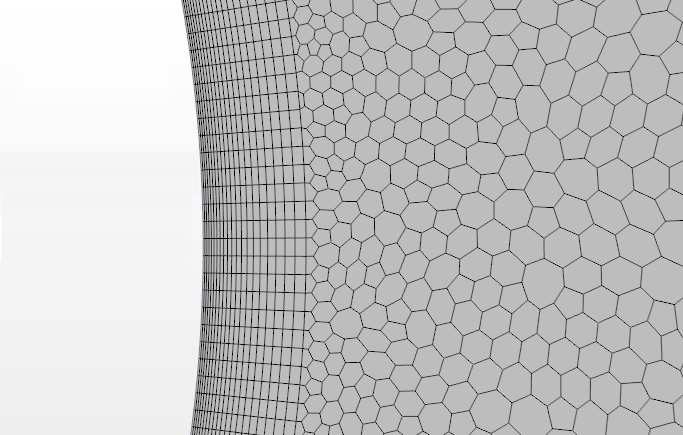}
		\caption{Expanded view around cylinder surface}
		\label{ellipse-S_L=1_mesh1}
	\end{subfigure}
	\begin{subfigure}{0.45\linewidth}
		\includegraphics[width=\linewidth]{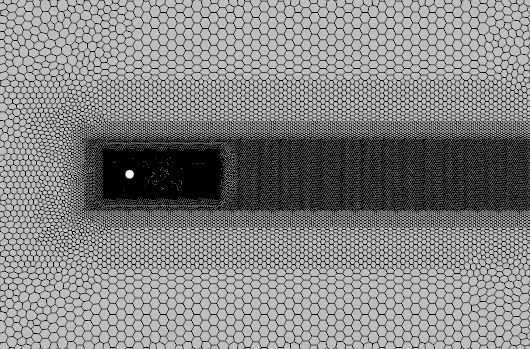}
		\caption{Overall view}
		\label{ellipse-S_L=1_mesh2}
	\end{subfigure}
	\caption{Computational domain showing (a) an expanded view of the mesh in the vicinity of the wall of the cylinder and (b) an overall view of the entire mesh in the computational domain.}
	\label{ellipse-S_L=1_mesh}
\end{figure}

For flow passing stationary circular cylinder, a study of the current numerical grid's dependence on mesh resolution was undertaken.
Table~\ref{ValidationRe100} summarizes the data of this grid-dependence investigation. 
It can be seen that the relative differences of $C_{L, max}$ and $S_t$ between mesh 1 to mesh 2 are considerable, but decrease to 0.29\% and 0.59\% as the mesh is refined to mesh 3 (fine) and mesh 4 (very fine).
Additionally, both $C_{L, max}$ and $S_t$ agree well with other numerical results summarized by Lu {\it et al.} \cite{Lu2011} and Zhang {\it et al.} \cite{zhang_li_ye_jiang_2015}.
In this case, mesh~3 (consisting of 66708 cells) is utilized for all simulations in this work considering it presents the optimum combination of numerical accuracy and computational cost.
Figure ~\ref{ellipse-S_L=1_mesh} shows the overall and extended view of the computational domain corresponding to mesh 3.

\begin{center}
	\begin{table}[H]
		\caption{Comparison between present work (with different mesh qualities) and other reported results \cite{Lu2011, zhang_li_ye_jiang_2015} with respect to the amplitude of the lift coefficient $C_L^{max}$ and Strouhal number ($St$) for flow passing stationary circular cylinder at $Re$ = 100.}
		\centering
		\begin{tabular}{  c c c c  c  c  c  c  c  }
			\hline
		    & Cells number & $C_{L,max}$ & $S_t$  \\
			\hline
			Mesh 1 & 19145 & 0.294 & 0.1729   \\
			Mesh 2 & 42035 & 0.319 & 0.1702   \\
			Mesh 3 & 66708 & 0.341 & 0.1692   \\
			Mesh 4 & 75525 & 0.342 & 0.1691   \\
			Lu {\it et al.} \cite{Lu2011} & \ & 0.340 & 0.1670   \\
			Zhang {\it et al.} \cite{zhang_li_ye_jiang_2015} & \ & 0.340 & 0.1660   \\ \hline
		\end{tabular}
		\label{ValidationRe100}
	\end{table}
\end{center}

\begin{figure}[t]
	\centering
	\begin{subfigure}[b]{0.495\linewidth}
		\includegraphics[width=\linewidth]{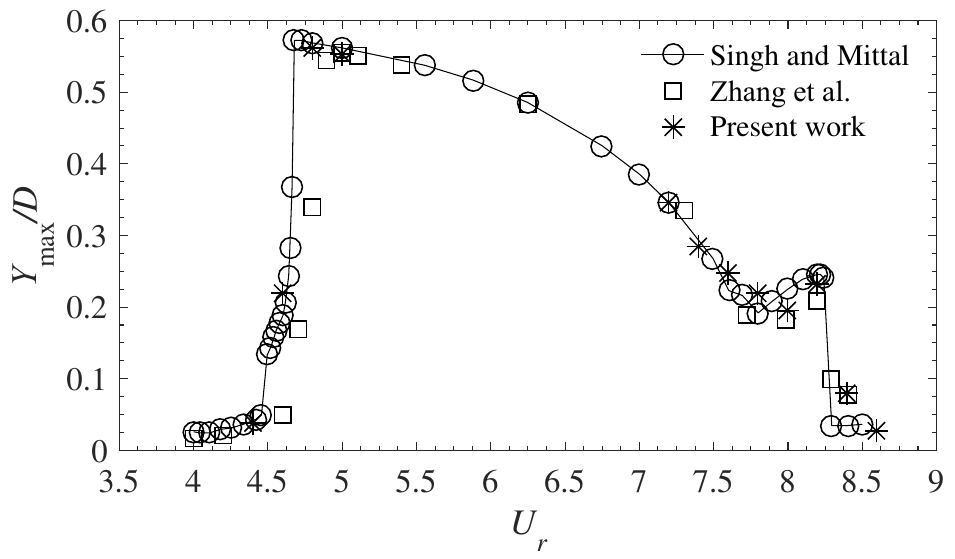}
		\caption{}
		\label{Model_verification1}
	\end{subfigure}
	\begin{subfigure}[b]{0.495\linewidth}
		\includegraphics[width=\linewidth]{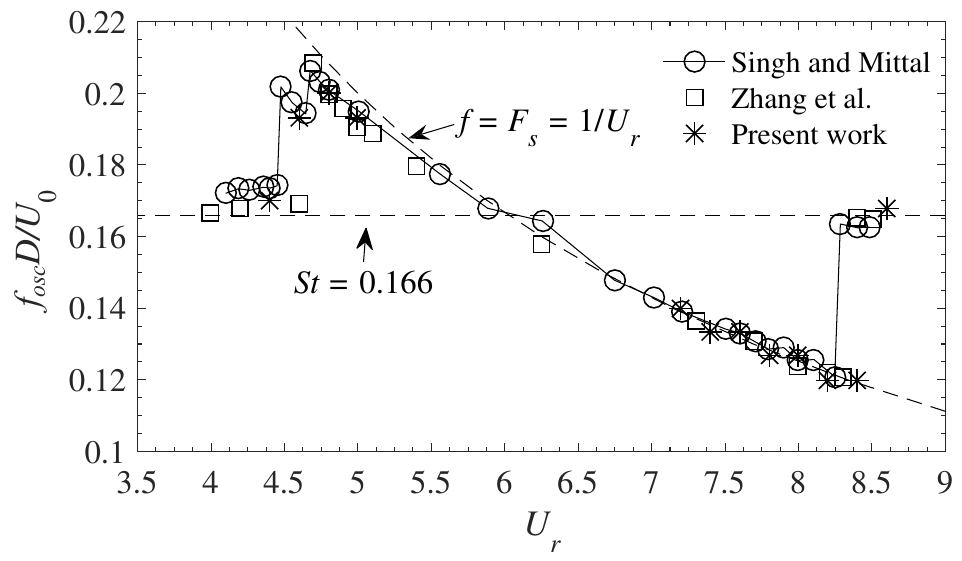}
		\caption{}
		\label{Model_verification2}
	\end{subfigure}
	\caption{Envelope of transverse (a) maximum amplitude and (b) oscillation frequency as a function of reduced velocity $U_r$ for a circular cylinder at $(Re,m^*)=(100,10)$. Comparison is between present results and those of Zhang {\em et al.}~\cite{zhang_li_ye_jiang_2015} and Singh and Mittal \cite{SINGH20051085}.}
	\label{CFDModel_verification}
\end{figure}

Prior to the study of the concerned configurations in this paper, the validation of the current fluid-structure coupling model and its implementation is required. The validation will be based on a VIV configuration submerged in 2-D  laminar flow---flow passing circular cylinder which is elastically mounted and confined to move freely in the cross-flow ($y$-) direction. The physical parameters for this configuration are as follows: structural damping coefficient $\zeta = 0$, mass ratio $m^* = 10$, and Reynolds number $Re = 100$. The reduced velocity \(U_r\) is changed by varying the structural natural frequency $f_n$ (i.e., spring stiffness). Present results will be compared with other numerical works and comparisons are summarized in Figure ~\ref{CFDModel_verification}. This figure shows the variation of the normalized maximum transverse displacement $Y_{{\rm max}}/D$ as a function of the reduced velocity \(U_r\).
In detail, $Y_{\rm max}/D$ exhibits a rapid increase around \(U_r \approx 4.6\) and then decreases slowly as $U_r$ increases.
It can be observed that the predictions in this paper are in good agreement with other numerical results \cite{SINGH20051085, zhang_li_ye_jiang_2015}.
The fluid-structure interaction model used in this work has good accuracy and could provide high-fidelity results for flow dynamics analysis as well as flow field data for subsequent identification processes.

\section{Strategy 1: Direct identification of wake oscillator}

\begin{figure}[H]
  \centering
  \begin{subfigure}[b]{0.55\linewidth}
    \includegraphics[width=\linewidth]{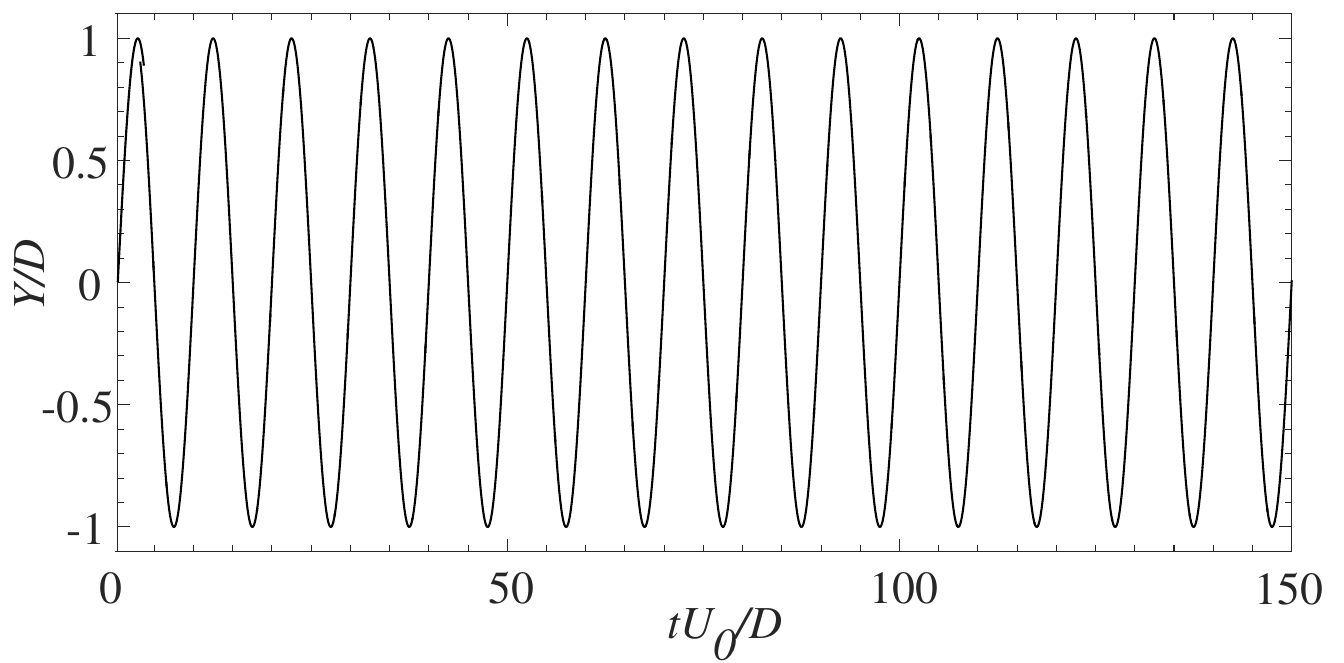}
	\caption{Designed cylinder motion. }
	\label{Y_Re100WithoutSFD_designedMotion_SameA1D_F0_1Hz}
  \end{subfigure}
  \begin{subfigure}[b]{0.55\linewidth}
    \includegraphics[width=\linewidth]{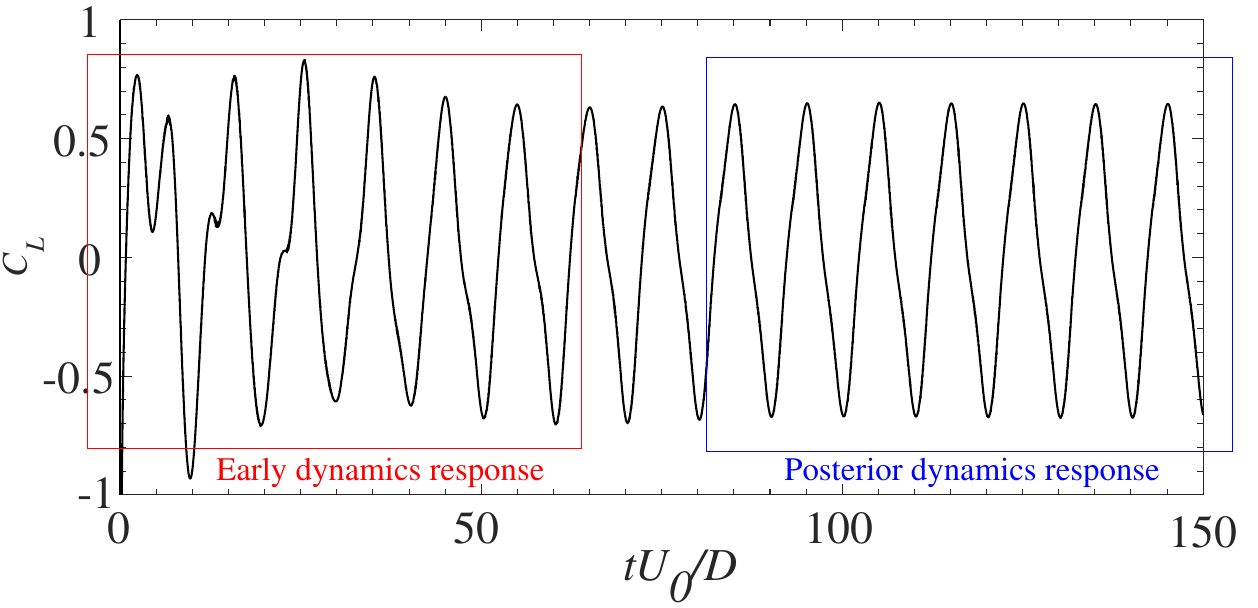}
	\caption{$C_L$ response. }
	\label{Re100WithoutSFD_designedMotion_SameA1D_F0_1Hz}
  \end{subfigure}

  \caption{Designed cylinder motion and $C_L$ response under the designed cylinder motion using FOM/CFD at $Re$ = 100. Early dynamics response and posterior dynamics response are marked with red and blue boxes, respectively.}
  \label{CFD_data_A1_F0.1}
\end{figure}

This section focuses on the direct identification of wake oscillators and the training data are obtained by the CFD calculation of moving circular cylinders with designed/forced oscillation.
The governing equations for the trajectory of the structure are shown in Eq. \ref{multiAeq}, where the variation of the oscillation amplitude and frequency are determined by the coefficients $ac$ as well as $oc$, respectively. When $ac$ and $oc$ are equal to 1, the structural trajectory is the periodic vibration with constant amplitude and frequency. In this work, the iso-amplitude and iso-frequency motion pattern at $Re$ = 100 is first selected to carry out the wake oscillator identification, and the dimensionless amplitude $A$ and the reduced vibration frequency $F_{osc}$ (=$f_{osc}D/U_0$) are set to be 1 and 0.1. The time history of the structural amplitude as well as the responded lift coefficients are shown in Fig. \ref{CFD_data_A1_F0.1}, where the time period after the lift reaches the equilibrium state is selected as the basis for the training.



\begin{linenomath}
\begin{equation}\label{multiAeq}
Y=\frac{A_{osc}}{ac^{( t-t_0 )}}\cdot sin( 2\pi f_{osc}\cdot oc^{( t-t_0 )}\cdot ( t-t_0 ) ),
\end{equation}
\end{linenomath}

\begin{figure}[H]
\centering
\centering\includegraphics[width=100mm]{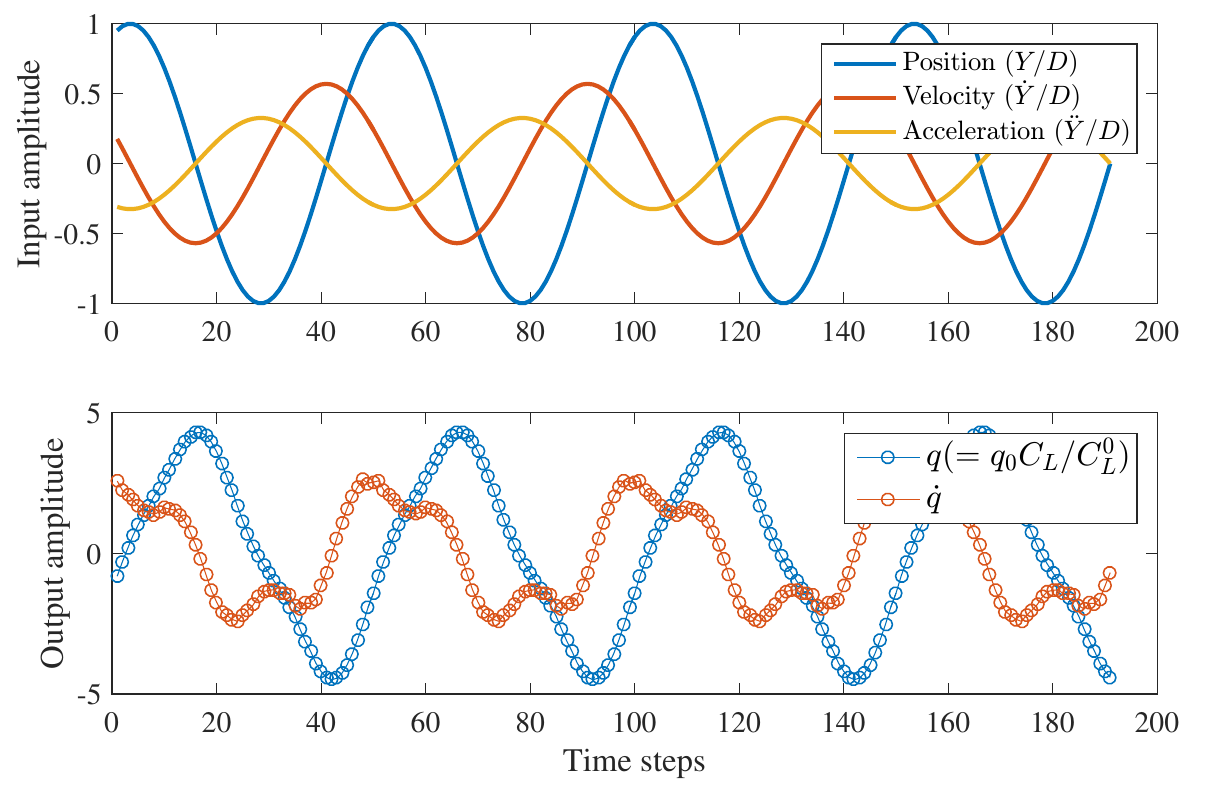}
\caption{Input and output data selected from the `posterior dynamics response' interval in Figure \ref{CFD_data_A1_F0.1} for identification.}
\label{chosenCLY_Re100WithoutSFD_designedMotion_SameA1D_F0_1Hz}
\end{figure}

We further processed the above data provided by the CFD calculations to obtain the inputs (cf. with $y$, $\dot{y}$, $\ddot{y}$) and outputs ($q$, $\dot{q}$) required for the identification of dynamic modeling (viz., Eq. \ref{wostatusequation}), which are presented in Fig. \ref{chosenCLY_Re100WithoutSFD_designedMotion_SameA1D_F0_1Hz}.
Regarding the determination of the initial status in identification process, its value does not need to be exactly consistent with the initial value of the output data (cf. with second panel of Fig. \ref{chosenCLY_Re100WithoutSFD_designedMotion_SameA1D_F0_1Hz}), and it is reasonable to limit it within a certain range. This is owing to the prediction results of the final identified dynamic mode cannot be exactly identical to the training data. 
This behavior will be further discussed later in the practical examples.

\begin{figure}[H]
\centering
\centering\includegraphics[width=100mm]{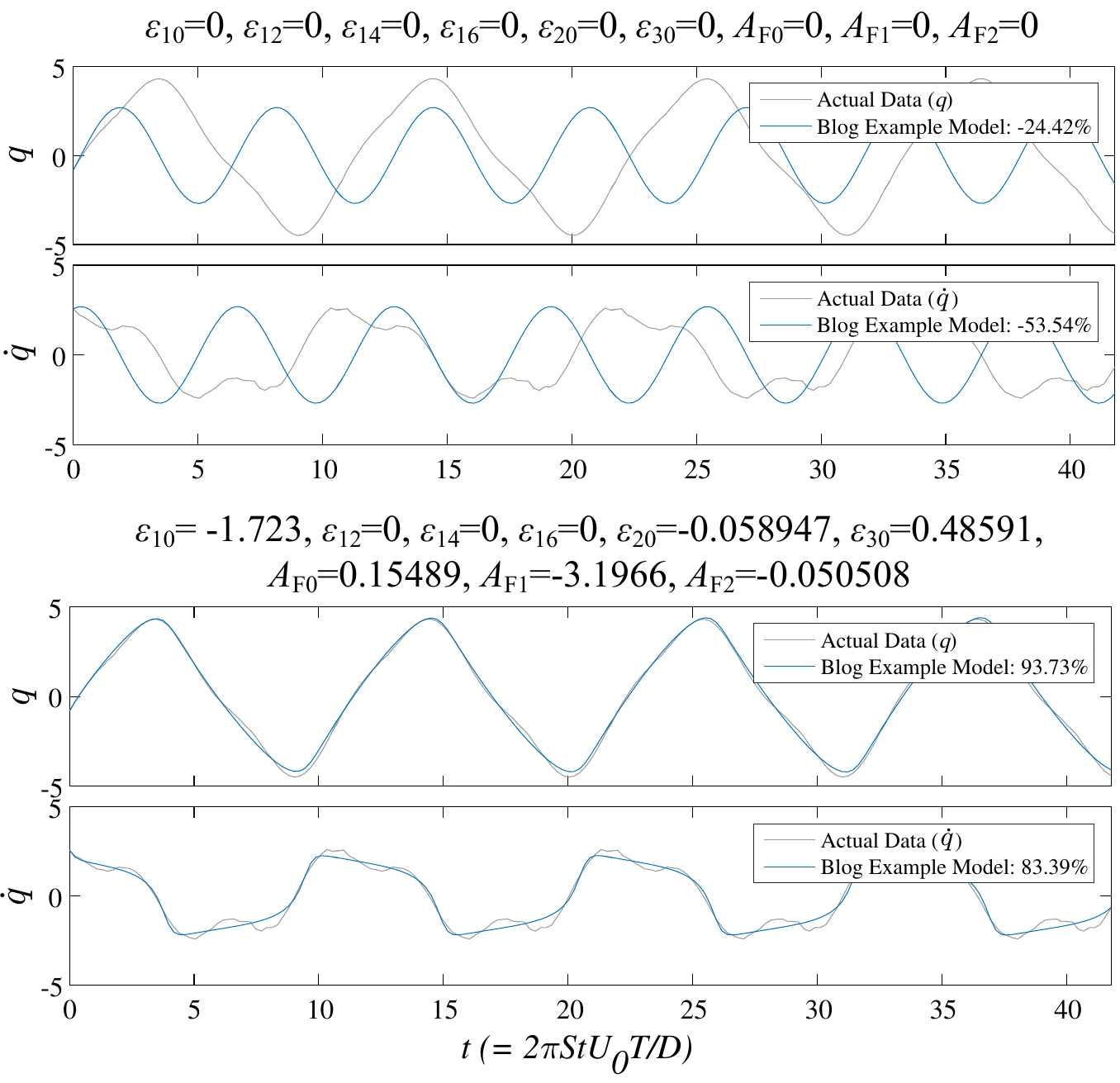}
\caption{Identified results of WO under the designed cylinder motion compared with those of using FOM/CFD for $Re$ = 100. $\varepsilon _{12}=\varepsilon _{14}=\varepsilon _{16}=0$.}
\label{nlgr_Re100WithoutSFD_designedMotion_SameA1D_F0_1Hz}
\end{figure}

For the given inputs (i.e., displacement, velocity, and acceleration information of the structure), 
the fit between the output (provided by the training data) and the prediction results for the original and identified models is exhibited in Figure \ref{nlgr_Re100WithoutSFD_designedMotion_SameA1D_F0_1Hz}.
Low-order nonlinear terms in equations \ref{wostatusequation} are sufficient to obtain an accurate representation of wake dynamics, so we set the coefficients $\varepsilon _{12}, \varepsilon _{14}, \varepsilon _{16}$ to 0 for the high-order terms.
The training data are given by the grey solid line, while the prediction results are given by the blue solid line. It can be observed that both for $q$ and $\dot{q}$, the prediction is significantly improved after identification, reaching 93.8\% and 83.4\%, respectively.
More specifically, with respect to the variation of $\dot{q}$, the prediction results are still unable to capture some of the subtle fluctuating features in the measured data.
It is expected that this is owing to high-order nonlinear damping terms are not considered in the present identification. 
However, the good fit level for displacements $q$ indicates that present identification strategy is sufficient for the concerned case herein.
As shown in Fig. \ref{nlgr_Re100WithoutSFD_designedMotion_SameA1D_F0_1Hz}, the parameters of the identified model at this attempt are finally represented as: ($\varepsilon _{10}, \varepsilon _{12}, \varepsilon _{14}, \varepsilon _{16}, \varepsilon _{20}, \varepsilon _{30}$) = (-1.723, 0, 0, 0, -0.05895, 0.4859) and ($A_{F0}, A_{F1}, A_{F1}$) = (0.1549, -3.197, -0.05051).

\begin{figure}[H]
\centering
\centering\includegraphics[width=170mm]{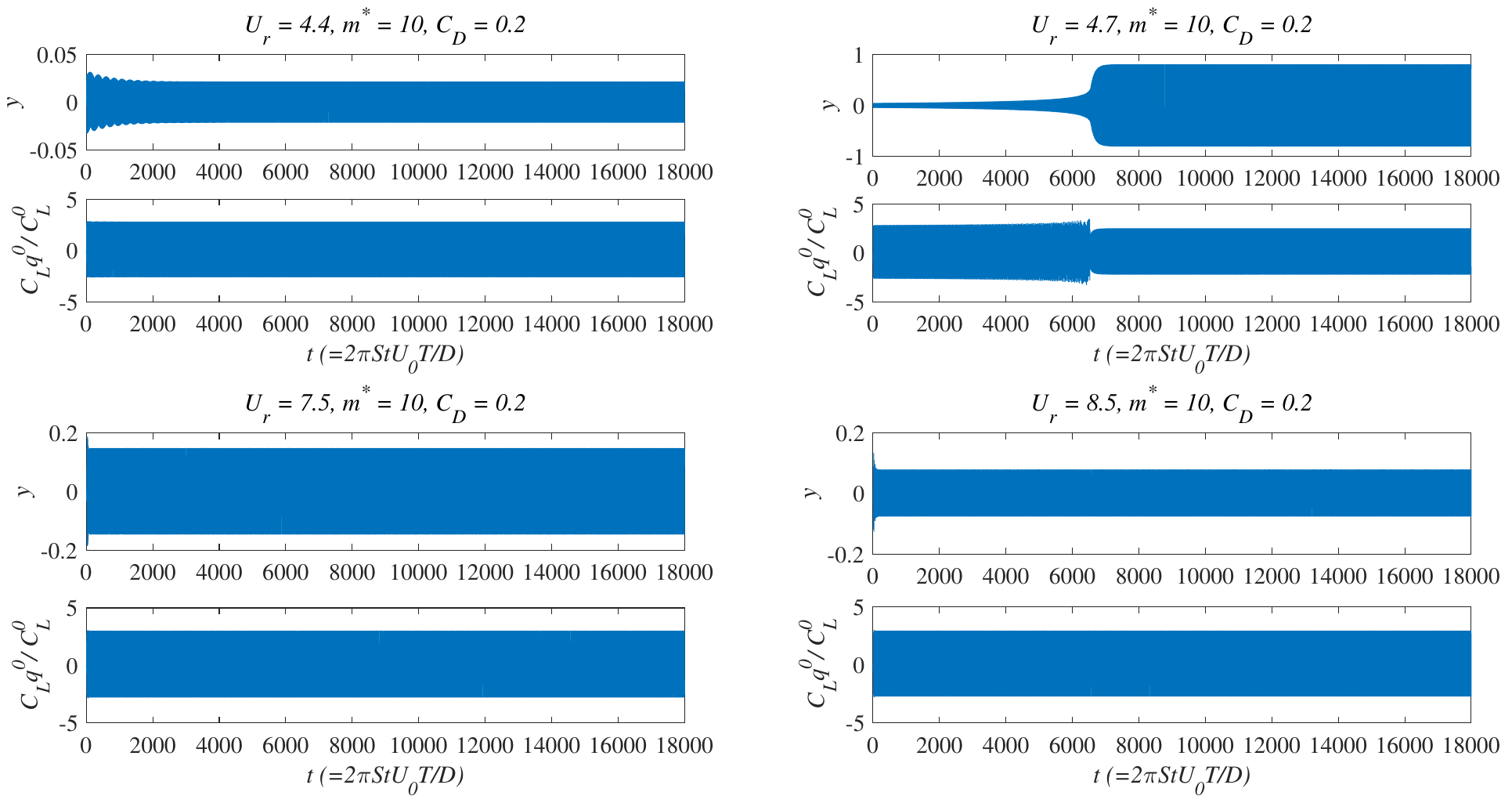}
\caption{Time history (0-18000 s) of $q$ and $y$ of VIV model coupling identified WO and structural equation for $(Re,m^*)$ = (100,10), with different $U_r$ of 4.4, 4.7, 7.5 and 8.5. }
\label{timehistory_Re100WithoutSFD_designedMotion_SameA1D_F0_1Hz}
\end{figure}

We incorporate the above-identified values into Eq. \ref{structuraleqa10} and couple it with the structural equations for VIV response prediction. 
$C_D$ is actually an estimate by the researchers on the effect of fluid damping, which is assumed to be 0.2 here.
To explore the accuracy of the present model, for the VIV response of the circular cylinder at ($Re,m^*$) of (100, 10), we choose several representative values for $U_r$ at (4.4, 4.7, 7.5, 8.5), carry out the VIV calculation, and exhibits its time-histories of correlated coefficients in Fig. \ref{timehistory_Re100WithoutSFD_designedMotion_SameA1D_F0_1Hz}.
As can be observed by the time history in the up panel of each subplot in Figure \ref{timehistory_Re100WithoutSFD_designedMotion_SameA1D_F0_1Hz}, the VIV system makes a magnitude jump from $U_r$ = 4.4 to 4.7 with respect to the displacement amplitudes, going from 0.02 to 0.81.
A more subtle phenomenon is that the predicted displacement response at $U_r$ = 4.7 undergoes a sudden increase in amplitude at $t$ = 6500. 
Such phenomena are also reported in past results obtained through CFD calculations \cite{prasanth_mittal_2008, CHENG2022113197}, originating from sudden destabilization of VIV system when the structural amplitude slowly reaches a certain threshold.
Moreover, the maximum structural amplitudes are 0.15 and 0.05 at $U_r$ = 7.5 and 8.5, respectively, exhibiting a decreasing trend with increasing $U_r$.

\begin{figure}[H]
\centering
\centering\includegraphics[width=100mm]{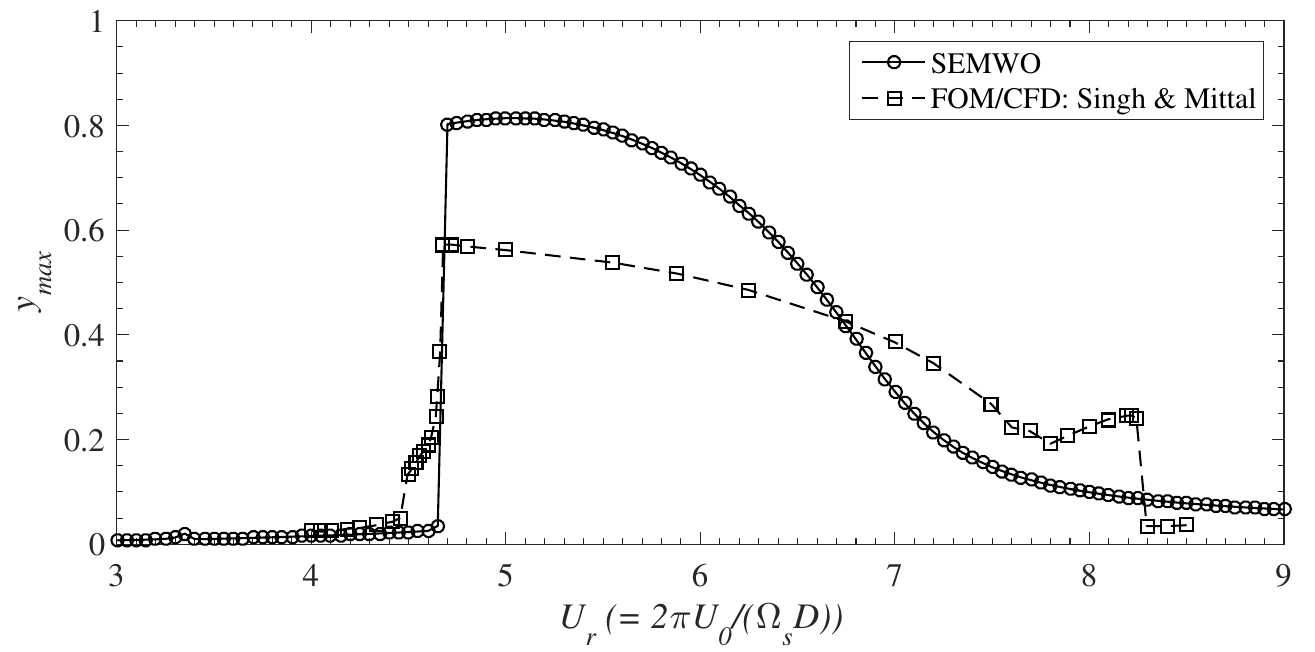}
\caption{Prediction of VIV maximum amplitude $y_{max}$ as a function of $U_r$ at $(Re,m^*)$ = (100,10), using SEMWO and FOM/CFD \cite{SINGH20051085}. }
\label{YmaxUr_Re100WithoutSFD_designedMotion_SameA1D_F0_1Hz}
\end{figure}

Furthermore, the maximum structural amplitudes $y_{max}$ as a function of reduced velocity $U_r$ for present concerned configuration (i.e., VIV system consisting of an elastically-mounted circular cylinder at ($Re, m^*$) = (100, 10)) are exhibited in Fig. \ref{YmaxUr_Re100WithoutSFD_designedMotion_SameA1D_F0_1Hz}, in which the present predictions are compared with the high-fidelity calculation results \cite{SINGH20051085}.
Similar to the amplitude envelope provided by CFD method, the present obtained model also exhibits a sudden increase at $U_r \approx$ 4.6, and then decreases slowly. The VIV response jumped out of the lock-in range around $U_r$ = 8, which agrees well between the high-fidelity CFD method and the identified wake oscillator model.
This ability to predict the instant collapse of structural stability (i.e., sudden amplification of structural amplitude with increasing reduced velocity) is not achievable with other prediction models (based on wake oscillators) proposed in past correlated works \cite{OGINK20105452,CHEN2022103530,KURUSHINA2018547}.

\begin{figure}[H]
\centering
\centering\includegraphics[width=170mm]{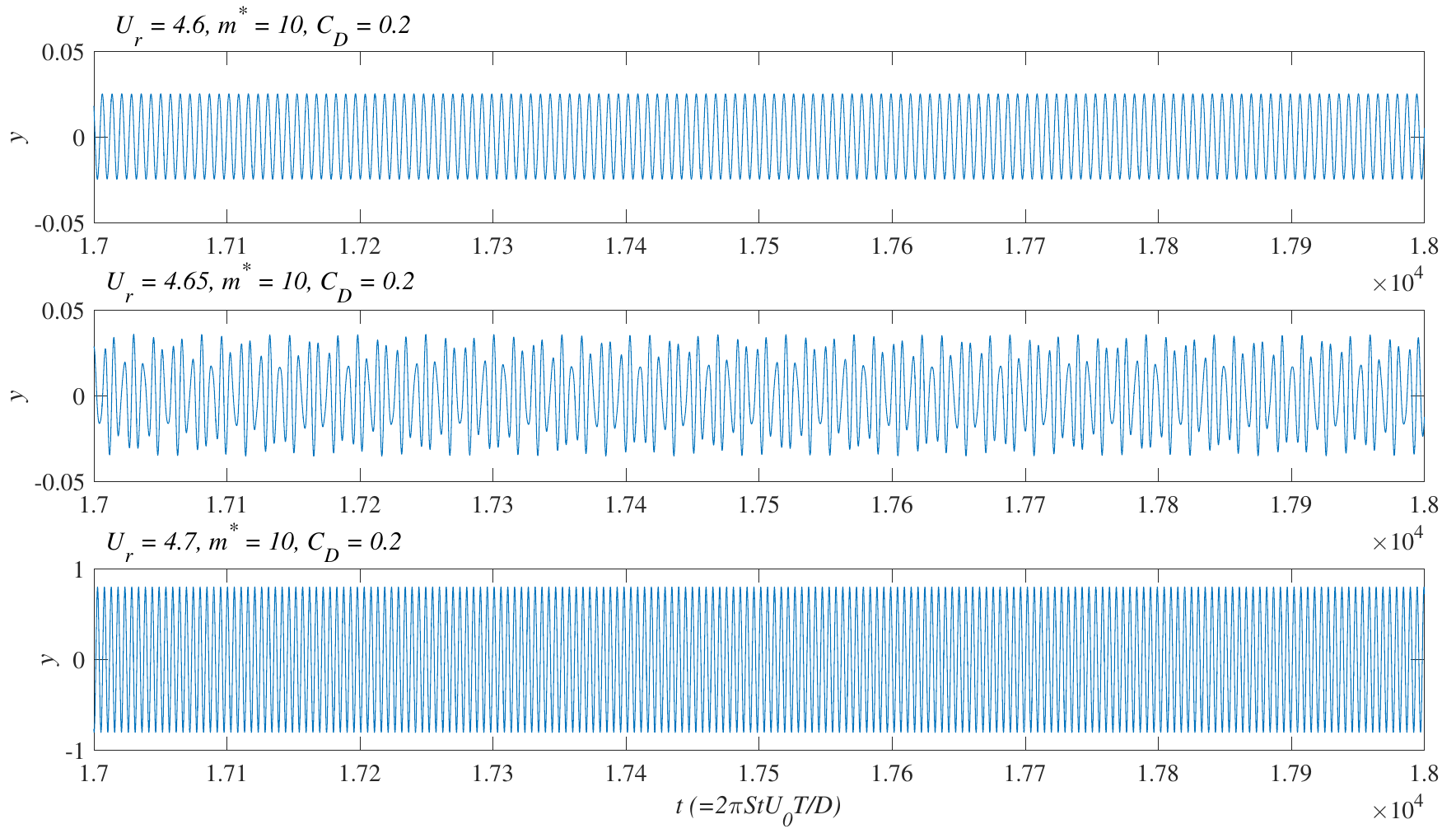}
\caption{Time history (17000-18000 s) of $y$ of semi-empirical VIV model coupling identified WO and structural equation (SEMWO) for $(Re,m^*)$ = (100,10), with different $U_r$ of 4.6, 4.65 and 4.7. }
\label{smalltimehistory_Re100WithoutSFD_designedMotion_SameA1D_F0_1Hz}
\end{figure}

In order to further explore the present model's ability for predicting the VIV response and capturing nonlinear features, we select the $U_r$ values close to the amplitude leap location and depict the corresponding amplitude time-histories after the system reaches equilibrium status.
As indicated in the first panel of Fig. \ref{smalltimehistory_Re100WithoutSFD_designedMotion_SameA1D_F0_1Hz}, the system is still located in the desynchronization regime at $U_r$ = 4.60, accompanied by a small structural displacement. As $U_r$ increases to 4.65, the amplitude evolution displays the `beating' phenomenon \cite{prasanth_mittal_2008}, which originates from the balanced competition between the structure mode and the wake modes \cite{ChengEtal2022,li_lyu_kou_zhang_2019}.
The successful capture of this behavior through wake oscillators was not achievable in past works.
Finally, the system jumps into the lock-in range when $U_r$ reaches 4.7. 
Generally, the present identified semi-empirical model is able to successfully represent the development of the VIV system spanning from desynchronization, lock-in, and desynchronization again, and capture some of the more microscopic dynamics. 

\begin{figure}[H]
\centering
\centering\includegraphics[width=100mm]{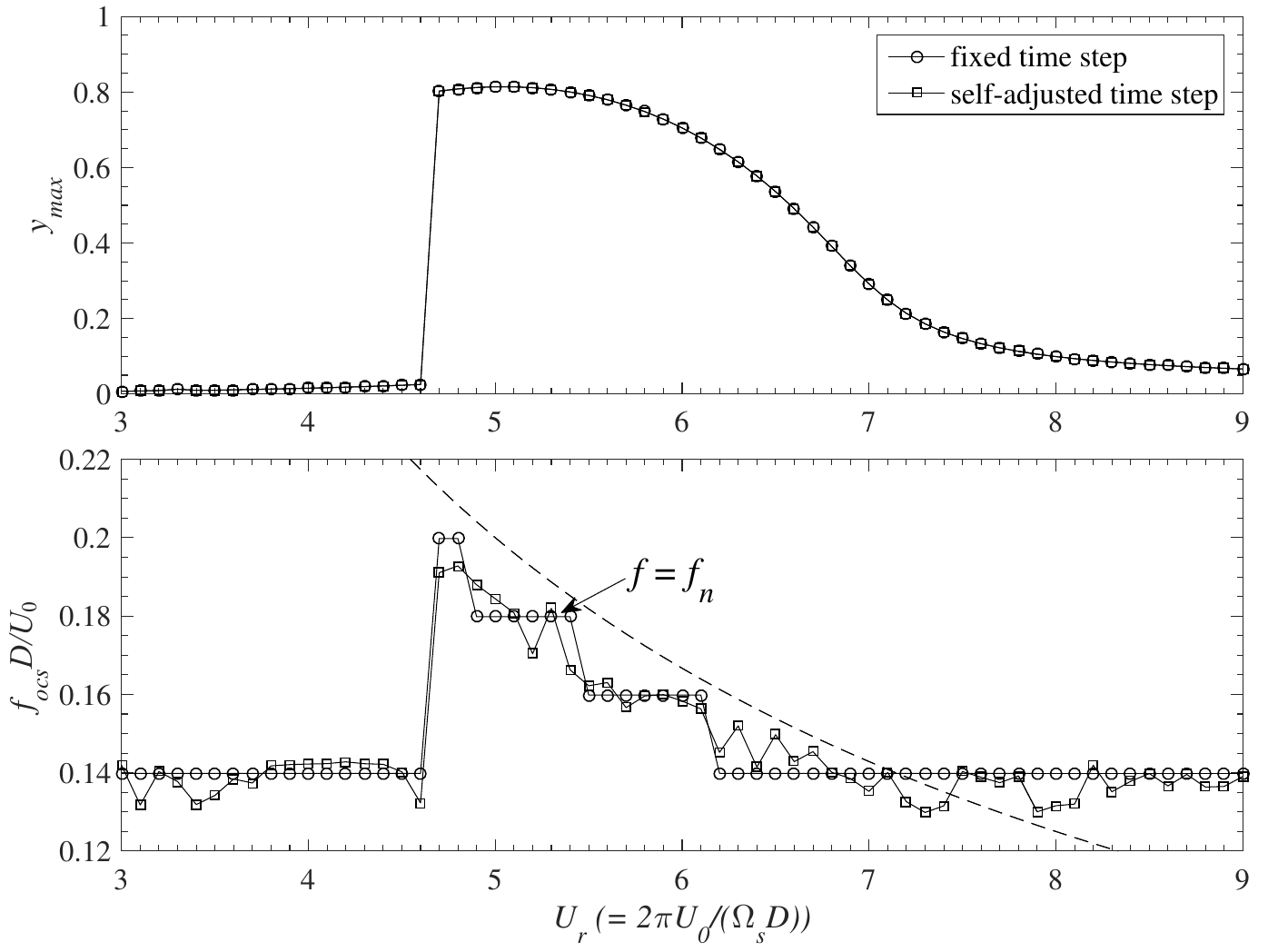}
\caption{Prediction of normalized maximum amplitude $y_{max}$ and oscillation frequency $f_{osc}D/U_0$ as a function of $U_r$ at $(Re,m^*)$ = (100,10), with fixed time step of 0.05 and self-adjusted time step during ODE.}
\label{YmaxFoscUr_CD0_2_tvt_Re100WithoutSFD_designedMotion_SameA1D_F0_1Hz}
\end{figure}

Fig. \ref{YmaxFoscUr_CD0_2_tvt_Re100WithoutSFD_designedMotion_SameA1D_F0_1Hz} exhibits the predictions obtained by two methods of determining the time steps (for the ODE calculation of obtained models). 
One applies the fixed time step and another uses the self-adjusting time step. Although the predictions of the two calculations show a completely identical performance in predicting the variation of the maximum amplitude, the frequency envelope with time-step variations exhibits more fluctuations. However, the predicted vibration frequency of the structure for both cases basically remains close to the natural frequency during the variation of $U_r$. 

\begin{figure}[H]
\centering
\centering\includegraphics[width=100mm]{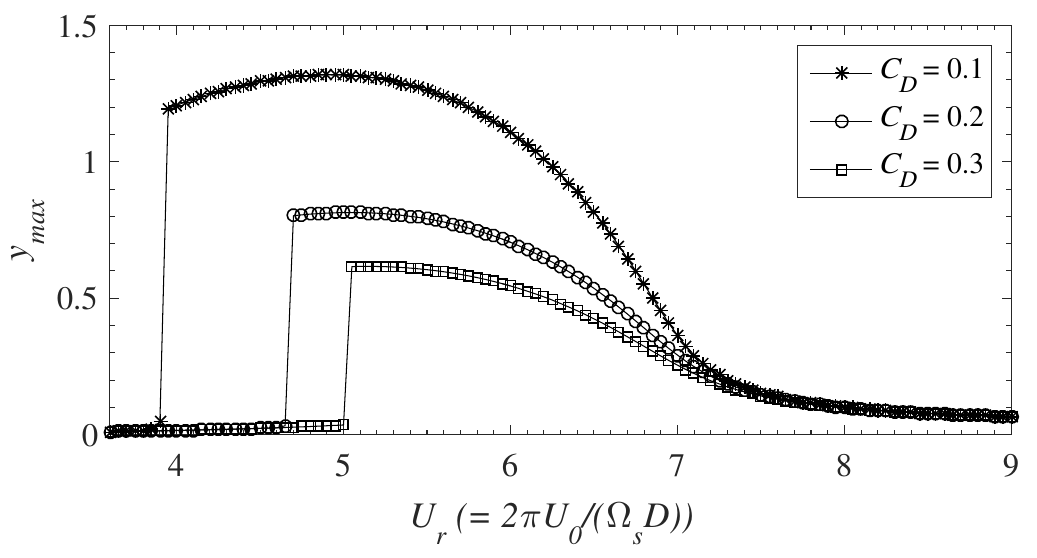}
\caption{Prediction of maximum amplitude $y_{max}$ as a function of $U_r$ at $(Re,m^*)$ = (100,10), with $C_D$ of 0.1, 0.2 and 0.3.}
\label{YmaxFoscUr_CD_Re100WithoutSFD_designedMotion_SameA1D_F0_1Hz}
\end{figure}

$\gamma$ in structural equation \ref{structuraleqa6} is a term characterizing the damping effect of the fluid, and has also been regarded as a semi-empirical parameter that had been continuously tuned by past researchers for good prediction \cite{OGINK20105452, gao2018novel}. $\gamma$ is more specifically determined by $C_D$, which has been set to 0.2 in the above works of this paper, and good predictions of amplitude envelope have been obtained in Fig. \ref{YmaxUr_Re100WithoutSFD_designedMotion_SameA1D_F0_1Hz}. To explore the impact of $\gamma$ in strategy 1, we show the results obtained for different $C_D$ in Fig. \ref{YmaxFoscUr_CD_Re100WithoutSFD_designedMotion_SameA1D_F0_1Hz}.
It can be observed that the assumed value of the fluid damping has a great influence on the onset $U_r$ as well as the maximum amplitude of the lock-in range. Overall, the smaller $C_D$ is accompanied by a more unstable VIV system.

\section{Strategy 2: Identification of VIV system}

\begin{figure}[H]
\centering
\centering\includegraphics[width=140mm]{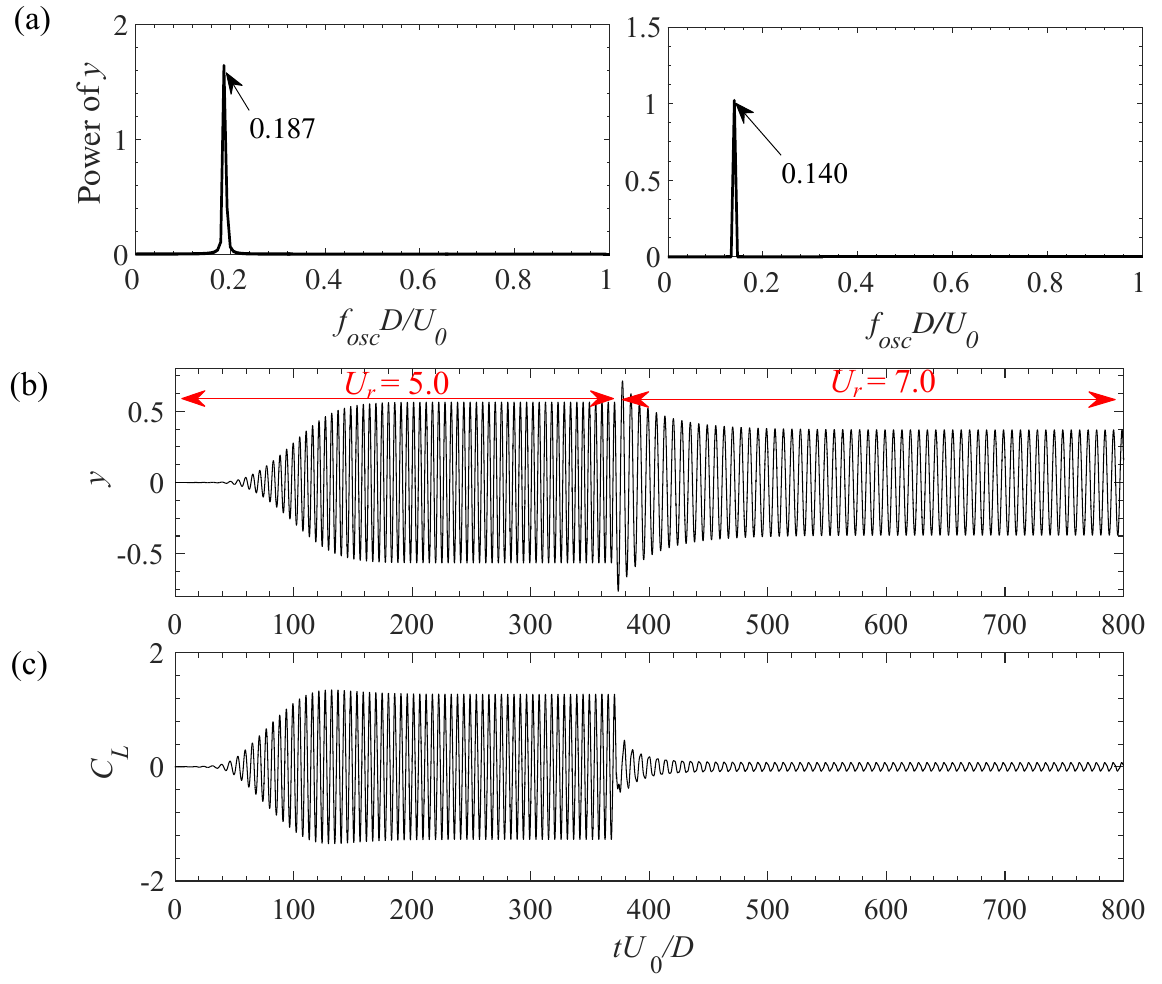}
\caption{The real-time history of normalized displacements $y/D$ and lift coefficients $C_L$ of VIV response (obtained by high-fidelity CFD method) consisting of the elastically-mounted circular cylinder with $U_r$ = 5 and 7 at ($Re,m^*$) = (100, 10).}
\label{VIV_Re100M10_Ur5and7_y_cl_andPSD}
\end{figure}

From the above discussion, it is implied that if the fluid equations (i.e., single wake oscillator) are identified in isolation, a semi-empirical term $\gamma$ remains in the structural equations, and the prediction of the VIV is still dependent on manual parameter tuning. Therefore, in the following, we will discuss strategy 2, i.e., integrating the fluid-solid coupled VIV governing equations as a whole into the identification framework.
In this section, the calculation for VIV response of the circular cylinder at $(Re, m^*)$ = (100, 10) is done for $U_r$ of 5 and 7, and the accompanying normalized structural displacements $y$ and lift coefficients $C_L$ are exhibited in Fig. \ref{VIV_Re100M10_Ur5and7_y_cl_andPSD} (b) and (c).
More specifically, $U_r$ = 5 and 7 correspond to the upper and lower branches of lock-in behavior, respectively. The frequency spectrum of $y$ at $U_r$ = 5 and 7 are exhibited in Fig. \ref{VIV_Re100M10_Ur5and7_y_cl_andPSD} (a).
In the following work, we will conduct the identification within strategy 2 based on the above VIV response data.

\subsubsection{$U_r$ = 5}
\label{di_VIVUr5}

\begin{figure}[H]
\centering
\centering\includegraphics[width=120mm]{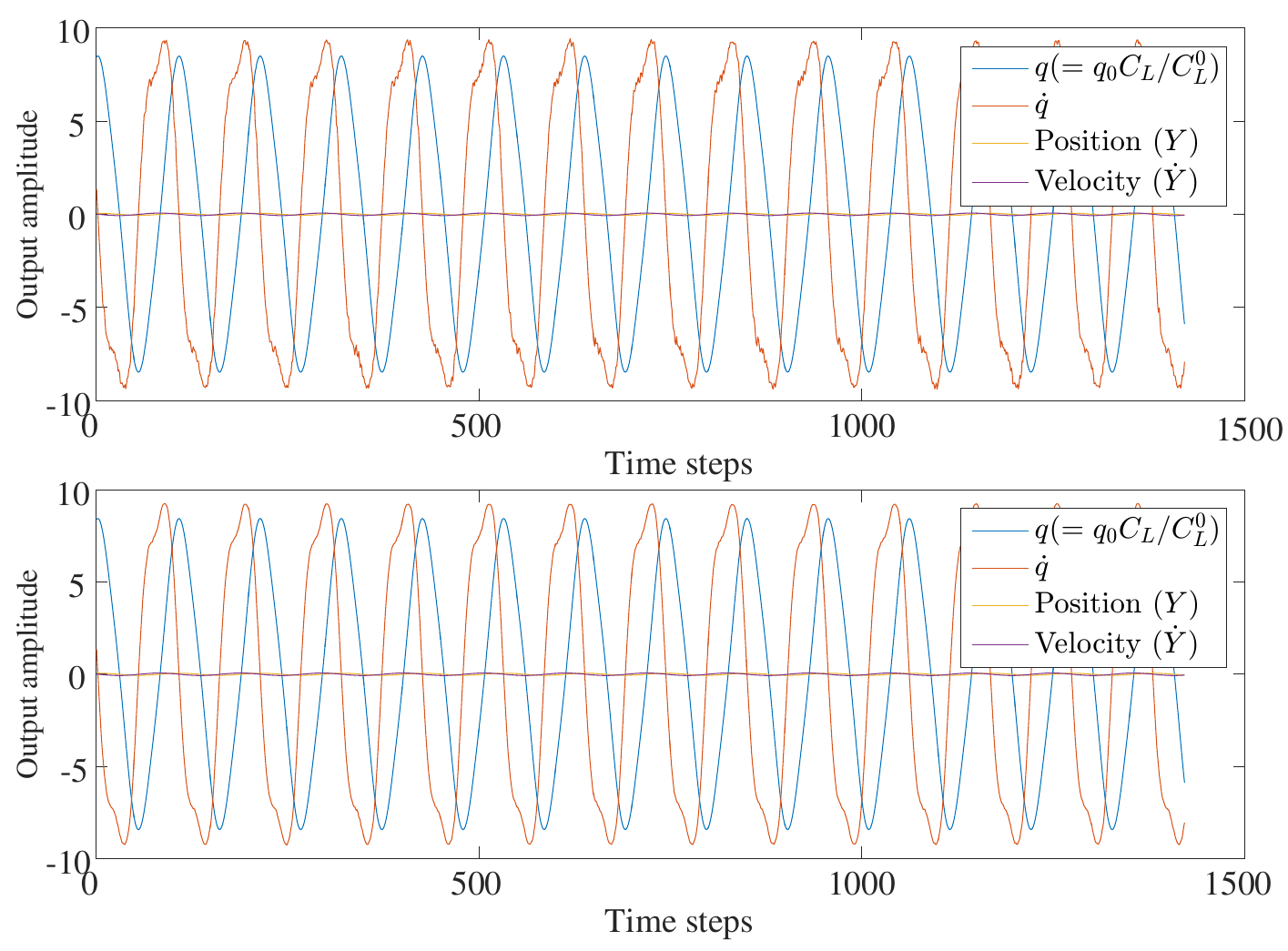}
\caption{(a) Original data obtained by CFD calculation, and (b) Smoothed data for the following training process for the real-time VIV response with respect to an elastically-mounted circular cylinder at ($Re, m^*, U_r$) = (100, 10, 5.0). The data include $q$ (normalzied lift coefficients), $\dot{q}$, $Y$ (displacements), and $\dot{Y}$ (velocity).}
\label{VIV_dimension_Re100M10_Ur5_y_cl_tobetrained}
\end{figure}

The original training data (prepared for the dynamical model \ref{VIVstatusequation}) of $q$, $\dot{q}$, $y$, and $\dot{y}$ are shown in the upper panel of Fig. \ref{VIV_dimension_Re100M10_Ur5_y_cl_tobetrained}. A perusal of the data indicates that original data involves considerable noise components, especially for time histories of $q$. These noise components come from numerical fluctuations in the CFD calculations and not primarily physic.
Obviously, these noise components will affect the identification process. In this case, we first smooth the original training data, and the smoothed training data are exhibited in the bottom panel of Fig. \ref{VIV_dimension_Re100M10_Ur5_y_cl_tobetrained}
Furthermore, when carrying out the identification of strategy 2, 
the damping term $\gamma$ in structural equation \ref{structuraleqa6}, which represents the damping effect from the fluid phase, could be completely replaced by the nonlinear features of the identified fluid equation \ref{structuraleqa10}, and thus removed.

\begin{figure}[H]
\centering
\centering\includegraphics[width=150mm]{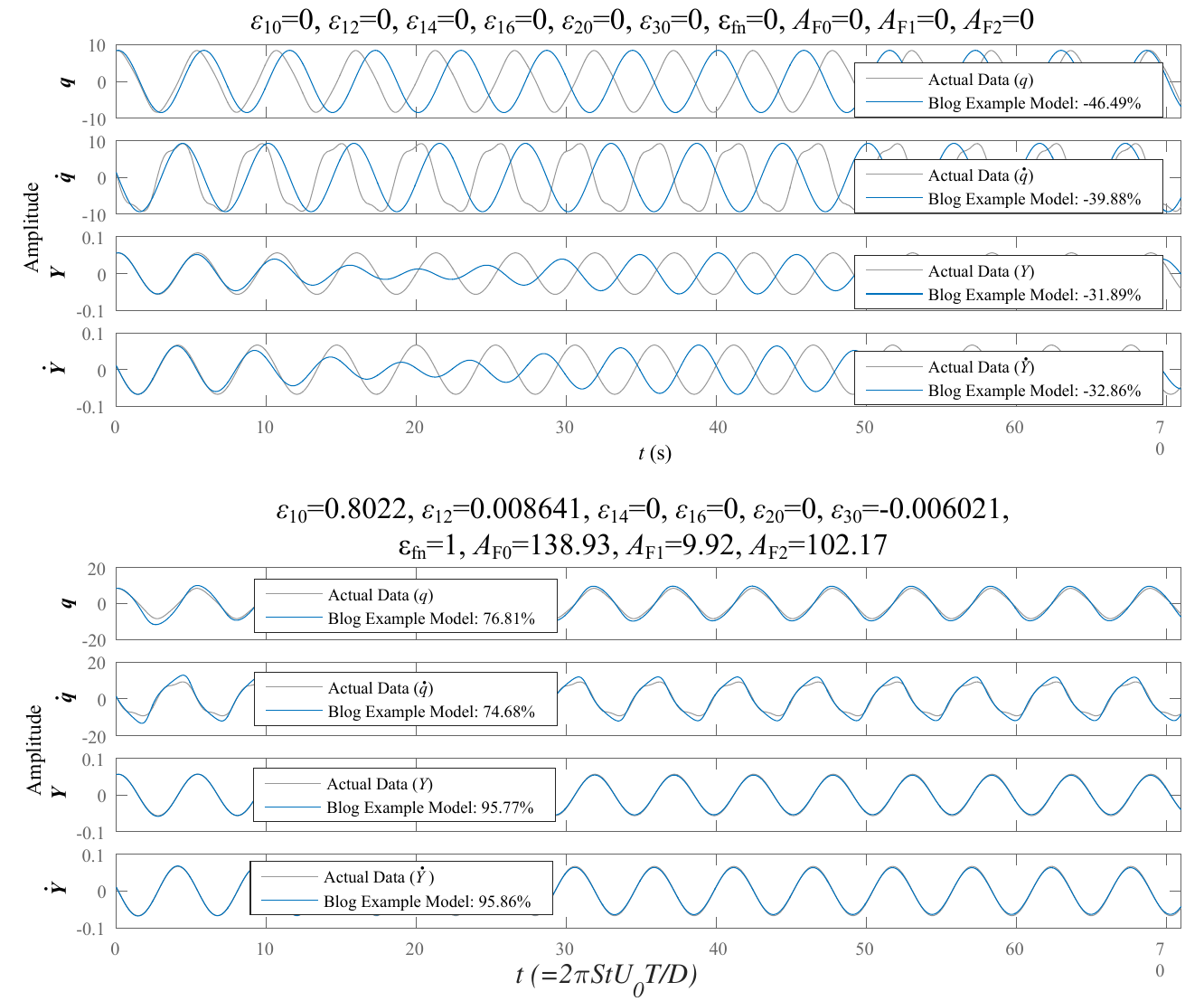}
\caption{Prediction of identified (semi-empirical) VIV control equations, in comparison to original training data provided by FOM/CFD at $(Re,m^*,U_r)$ = (100, 10, 5.0). $\varepsilon _{14}=\varepsilon _{16}=\varepsilon _{20}=0$.}
\label{nlgr2_WO_VIV_dimension_Re100_m10_Ur5_CD0}
\end{figure}

\begin{figure}[H]
\centering
\centering\includegraphics[width=110mm]{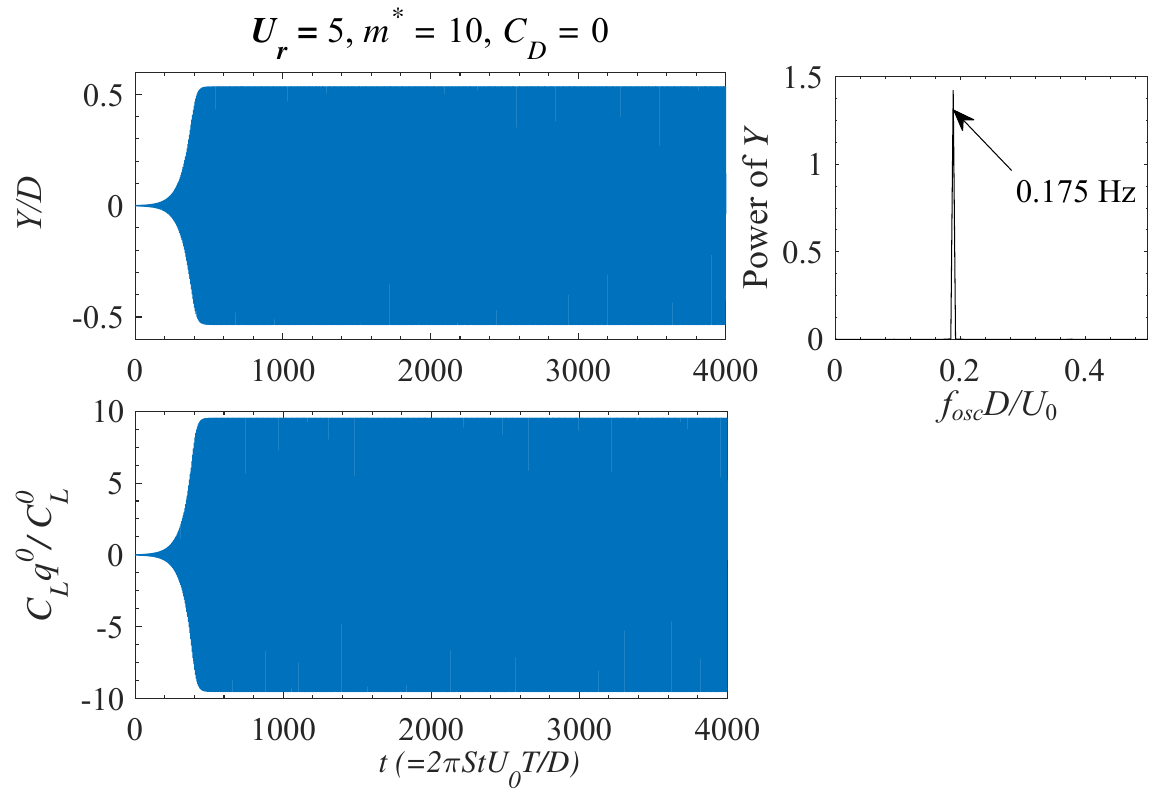}
\caption{Time history (0-4000 s) of $q$ and $y$ for the system response predicted by the coupled VIV equation (integrating wake oscillators and structural control equations) identified via strategy 2.
$(Re,m^*,U_r)$ = (100,10,5.0).}
\label{realtimeandPSD_nlgr2_WO_VIV_dimension_Re100_m10_Ur5_CD0}
\end{figure}

We start by carrying out the first recognition attempt, $\varepsilon_{14}$ = 0, $\varepsilon_{16}$ = 0, $\varepsilon_{20}$ = 0, $\varepsilon_{40}$ = 0, $\varepsilon_{50}$ = 0.
For the given inputs (i.e., displacement, velocity, and acceleration information of the structure), the change of the fits between the prediction results and the output (provided by the training data) is exhibited in Figure \ref{nlgr2_WO_VIV_dimension_Re100_m10_Ur5_CD0}.
It is obvious that the identified model could accurately match the time-histories development of the properties, although it is difficult to capture features of some local variations. 
We apply the identified model for the prediction at ($Re,m^*,U_r$) = (100, 10, 5), and plot the predicted data in Fig. \ref{realtimeandPSD_nlgr2_WO_VIV_dimension_Re100_m10_Ur5_CD0}.
For both structural displacements and dynamic coefficients of the VIV system response, the identified model provides good prediction accuracy.

\subsubsection{$U_r$ = 7}

\begin{figure}[H]
\centering
\centering\includegraphics[width=120mm]{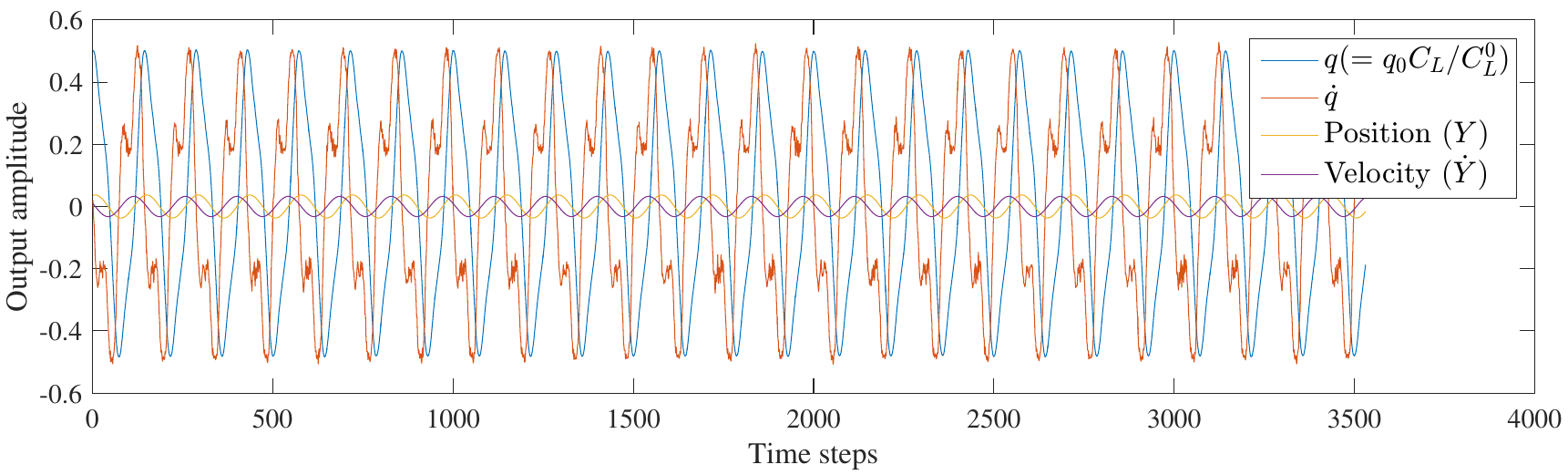}
\caption{Smoothed training data (obtained from CFD method) for the real-time VIV response with respect to an elastically-mounted circular cylinder at ($Re, m^*, U_r$) = (100, 10, 7.0). The data include $q$ (normalzied lift coefficients), $\dot{q}$, $Y$ (displacements), and $\dot{Y}$ (velocity).}
\label{VIV_dimension_Re100M10_Ur7_y_cl_tobetrained}
\end{figure}

Sub-section \ref{di_VIVUr5} considers the identification for $U_r$ = 5 of the upper branch in VIV response at ($Re,m^*$) = (100, 10), and this sub-section will consider the situation of $U_r$ = 7 belonging to lower branch, and the smoothed training data are exhibited in Fig. \ref{VIV_dimension_Re100M10_Ur7_y_cl_tobetrained}.

\begin{figure}[H]
\centering
\centering\includegraphics[width=1.0\linewidth]{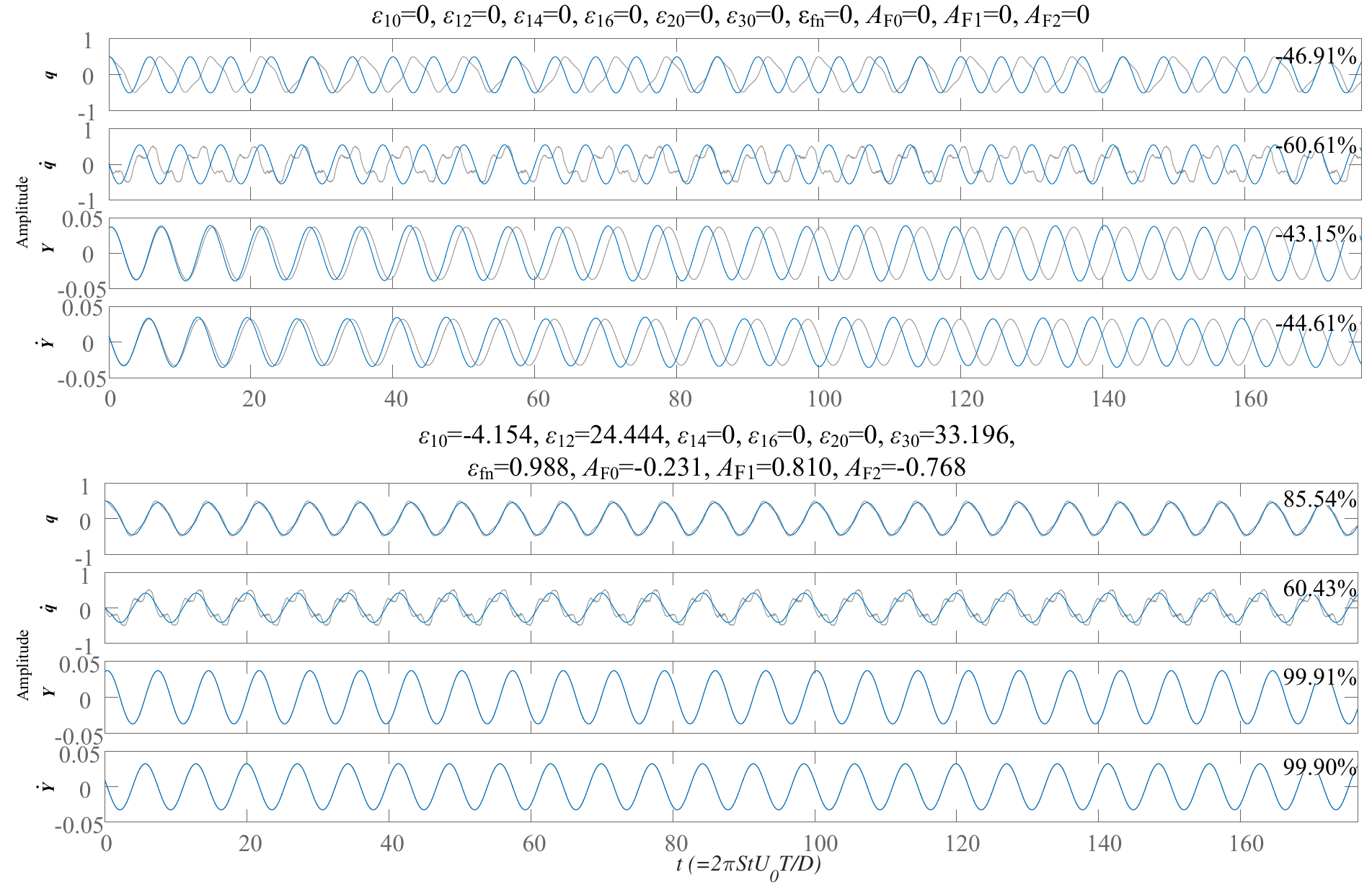}
\caption{Comparison of prediction results before and after model identification at $(Re,m^*,U_r)$ = (100, 10, 7.0). 
Predictions of (semi-empirical) VIV control equations based on WO are depicted with blue color, and original training data provided by FOM/CFD are marked with grey color.
$\varepsilon _{14}=\varepsilon _{16}=\varepsilon _{20}=0$.}
\label{nlgr_WO_VIV_dimension_Re100_m10_Ur7_CD0}
\end{figure}

\begin{figure}[H]
\centering
\centering\includegraphics[width=90mm]{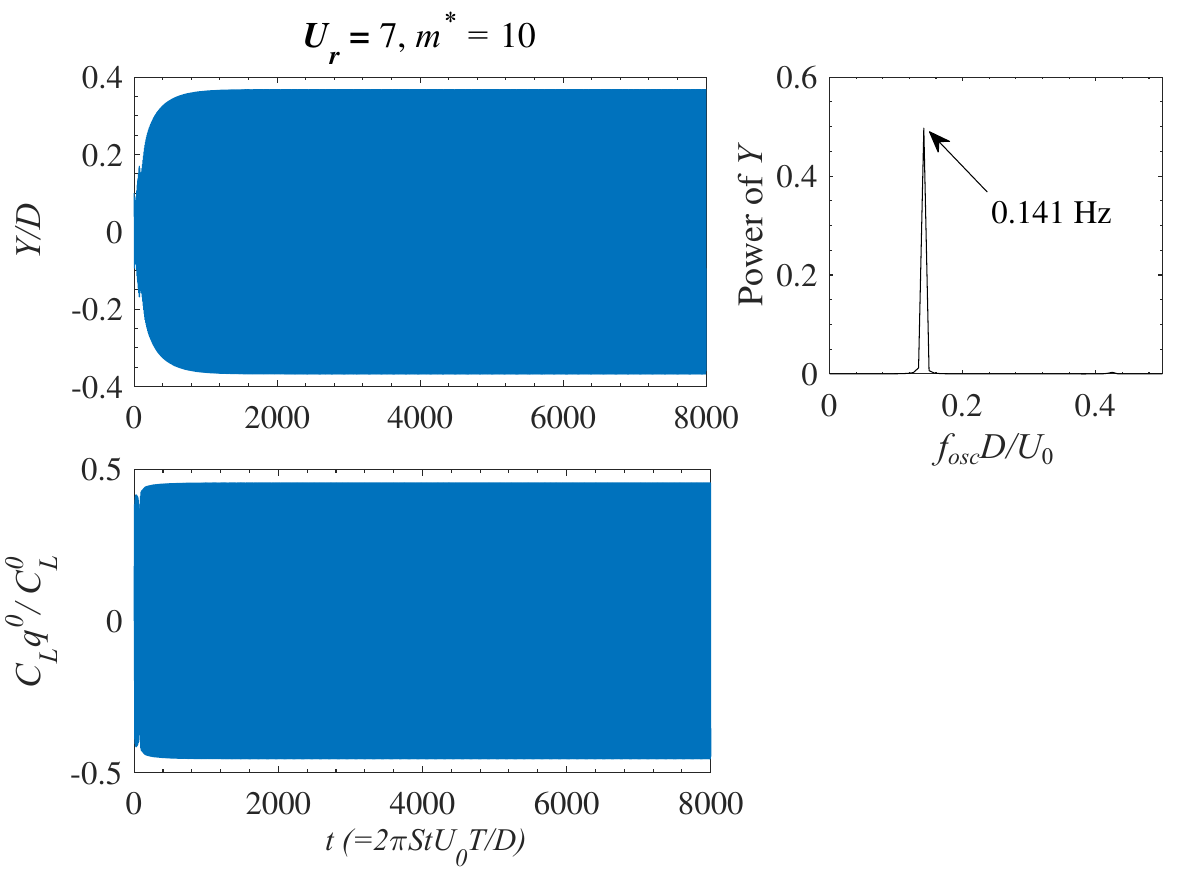}
\caption{Time history (0-8000 s) of $q$ and $y$ for the system response predicted by the coupled VIV equation (integrating wake oscillators and structural control equations) identified via strategy 2.
$(Re,m^*,U_r)$ = (100, 10, 7.0).}
\label{realtimeandPSD_nlgr_WO_VIV_dimension_Re100_m10_Ur7_CD0}
\end{figure}

Following up on the above identification steps of $U_r$ = 5.0, we continue processing the correlated data at $U_r$ = 7.0. The fits between predictions and training data are displayed in Fig. 
\ref{nlgr_WO_VIV_dimension_Re100_m10_Ur7_CD0}. Similar to the situation exhibited for $U_r$ = 5.0 in Fig. \ref{nlgr_WO_VIV_dimension_Re100_m10_Ur7_CD0}, the fit level with respect to the structural moving information is almost perfect, with the value up to 99.9\% (cf. with the information presented in the bottom two panels). However, there is some error in the prediction of the lift coefficients.
The post-identified model is applied for the prediction of the system at $U_r$ = 7 and the results are displayed in Fig. \ref{realtimeandPSD_nlgr_WO_VIV_dimension_Re100_m10_Ur7_CD0}.
Similar to the prediction at $U_r$ = 5 in Fig. \ref{realtimeandPSD_nlgr2_WO_VIV_dimension_Re100_m10_Ur5_CD0}, the model obtained by strategy 2 is also very accurate in predicting the structural displacements as well as the dynamic coefficients at $U_r$ = 7, and the prediction of the vibration frequencies is also accurate.

It should be noted here that we have made certain restrictions on the parameters (range or fixed value) during the model identification process. If we relieve these restrictions, we could obtain other combinations of parameters that are also capable of giving good prediction results. However, it is noted that whether it is the grey box method used in this work or other SINDy \cite{Steven2016,Quade2018} as well as machine-learning methods \cite{BAI2022105266, Chizfahm2022}, the mechanism underlying the identification process is purely mathematically based. Therefore, the magnitude of obtained/predicted model parameters may be very significant but still lack physical meaning.
In this case, based on the physical context already known to researchers, a reasonable limit on the parameters allowed the recognition tool to find the optimal result within a certain range.

With respect to the cases introduced in this work, the prediction results at $U_r$ equal to 5 or 7 both demonstrate the feasibility and accuracy of the proposed strategy in predicting the VIV response. 
Future work will be carried out in two steps: (1) For a certain Reynolds number, e.g., 100 in this paper, more intensive $U_r$ calculations will be carried out to obtain the model parameter values $\varepsilon_{ij}$ corresponding to each $U_r$, and then a semi-empirical formulation of $\varepsilon$ with respect to $U_r$ is established. (2) Then we can arrange the Reynolds numbers from low to high (50-500,000), and select representative Reynolds numbers for identification based on the one-third octave rule. 
More high-order nonlinear terms inside equation \ref{wostatusequation} will be involved in the identification at high Reynolds number flows.
The effect of Reynolds numbers is thereby encapsulated in the above semi-empirical formulae for $\varepsilon_{ij}$ as well, so as to build up a unified framework of VIV prediction based on the general wake oscillator.

\begin{linenomath}
\begin{equation}\label{wostatusequation}
\begin{aligned}
\left( \begin{matrix}
	\varepsilon _{10}( Re,U_r)&		\varepsilon _{12}( Re,U_r)&		\varepsilon _{14}( Re,U_r)&		...\\
	\varepsilon _{30}( Re,U_r)&		\varepsilon _{32}( Re,U_r)&		\varepsilon _{34}( Re,U_r)&		...\\
	\varepsilon _{50}( Re,U_r)&		\varepsilon _{52}( Re,U_r)&		\varepsilon _{54}( Re,U_r)&		...\\
	\vdots&		\vdots&		\vdots&		\ddots\\
\end{matrix} \right) \,\,
\end{aligned}
\end{equation}
\end{linenomath}


\section{Conclusion}

In this paper, an effort has been made to introduce new strategies to develop a unified framework for vortex-induced vibration (VIV) prediction based on semi-empirical wake oscillators.
This framework's development is based on the Greybox nonlinear system identification method using high-fidelity CFD data and/or experimental results. 
We first prepare one general wake oscillator template with many internal low- to high-order damping terms to be identified. Two strategies are introduced in this work: 1. the individual identification of the wake oscillator and then coupling it to the structural control equations to predict the VIV response, and 2. the overall identification of the structural equations coupling with fluid wake oscillator.

In strategy 1, we first apply CFD method to conduct the detailed and time-consuming calculation for the flow dynamics system consisting of one moving circular cylinder with a designed motion strategy. The parameters inside the general wake oscillator are thereby identified based on the training data provided by the CFD method. The prediction exhibits good accuracy regardless of the lock-in range and structural amplitude.

However, the prediction of strategy 1 still relies on the adjustment of fluid-damping terms inside structural control equations, and thus, we also considered strategy 2. The high-fedity response of two VIV systems at ($Re, m^*, U_r$) = (100, 10, 5.0) and (100, 10, 7.0) are obtained and then used for the identification of overall coupled VIV models based on structural and flow wake oscillators. Present results indicate that the identified models could provide nearly conformance compared to the original training data (i.e., high-fidelity physical response).

This paper then addresses the ultimate goal (from the mathematical view) and the following works for the present proposed unified framework. More specifically, the coefficients correlated to each term inside wake oscillators will be identified at representative properties combo (viz., Reynolds number, mass ratio, and reduced velocity) for the VIV system. One fit function of damping coefficients varying with those properties will be obtained. Future work will be conducted following the above-mentioned ideas and could provide great convenience for researchers as well as engineering applications once this framework is established.
In addition to the VIV prediction of circular cylinders, the process of building this prediction framework could also be used for FIV analyses of other geometries, such as the galloping prediction of square cylinders and/or airfoils.

\section{Acknowledgements}

The first author is supported by Natural Sciences and Engineering Research Council of Canada (NSERC) and CF Energy. This work was made possible by the facilities of Aeroelasticity Group at Duke University, the Shared Hierarchical Academic Research Computing  (\href{http://www.sharcnet.ca}{SHARCNET}), and Compute/Calcul Canada. ~\\

\bibliography{Vanderpol_report_matlab}

\end{document}